\documentclass[10pt,a4paper]{article}
\usepackage[utf8]{inputenc}
\usepackage{amsmath}
\usepackage{amsfonts}
\usepackage{amssymb}
\usepackage{color}
\usepackage{graphicx}
\usepackage{float}
\usepackage[square]{natbib}
\usepackage{doi}
\hypersetup{hidelinks}
\usepackage[a4paper, total={6in, 8in}]{geometry}
\usepackage{tikz}
\usepackage[export]{adjustbox}
\usepackage{layouts}
\usepackage{psfrag}
\usetikzlibrary{shapes.geometric, arrows}
\newcommand*\diff{\mathop{}\!\mathrm{d}}
\usepackage[hang,flushmargin,symbol]{footmisc}

\begin{document}

\begin{center}
\textbf{\Large{Incorporating sufficient physical information into artificial neural networks: a guaranteed improvement via physics-based Rao-Blackwellization}}
~\\
~\\
~\\
Gian-Luca Geuken, J\"orn Mosler and Patrick Kurzeja\footnote{\noindent Corresponding author\\Email adress: patrick.kurzeja@tu-dortmund.de}
~\\
~\\
\textit{Institute of Mechanics, TU Dortmund, Leonhard-Euler-Str. 5, 44227 Dortmund, Germany}
~\\
~\\
\end{center}
Abstract:
The concept of Rao-Blackwellization is employed to improve predictions of artificial neural networks by physical information. The error norm and the proof of improvement are transferred from the original statistical concept to a deterministic one, using sufficient information on physics-based conditions. The proposed strategy is applied to material modeling and illustrated by examples of the identification of a yield function, elasto-plastic steel simulations, the identification of driving forces for quasi-brittle damage and rubber experiments. Sufficient physical information is employed, e.g., in the form of invariants, parameters of a minimization problem, dimensional analysis, isotropy and differentiability. It is proven how intuitive accretion of information can yield improvement if it is physically sufficient, but also how insufficient or superfluous information can cause impairment. Opportunities for the improvement of artificial neural networks are explored in terms of the training data set, the networks' structure and output filters. Even crude initial predictions are remarkably improved by reducing noise, overfitting and data requirements. 
~\\
~\\
Keywords: artificial neural networks, Rao-Blackwell, sufficiency, information, material modeling
\thispagestyle{empty}
\clearpage
\section{Introduction}
\subsection{Motivation: Connecting artificial neural networks and physical information}
The aim of this work is to incorporate sufficient physical information into an artificial neural network (ANN) by a statistical concept, namely, Rao-Blackwellization. It stands out by the possibility to provide guaranteed improvements even for crude estimators. Little attention has been paid to this concept of statistical modeling from the viewpoint of deterministic physical modeling, although the combination of physical knowledge and analytical or numerical methods has a rich tradition. The development of specialized finite elements, as one classical example, has often been driven by knowledge about the physical system of interest, e.g., reduced integration to avoid locking problems for nearly incompressible materials or shell structures with plane-stress states \citep{brezzi_1991,belytschko_2014}. Also multi-scale problems can benefit from efficient information transfer, e.g., see a physics-based adoption of the Rao-Blackwell concept to capillary effects in \citep{kurzeja2016}. The present focus of application is oriented towards ANN-based material modeling, because it benefits strongly from information-based improvement strategies.

More recently, the emerging field of data-based models and ANNs in particular are found to be enhanced by physical information. One of the first combinations of artificial neural networks and physical modeling is the work of \citet{forssell1997} for tank level modeling. An early work in the field of constitutive modeling of solid mechanical systems is from \citet{liang2008} for elastomeric foams. \citet{bhadeshia2009} stated that "it is not clear how a neural network trained on the output of physical models actually captures the principles of the physical models and more tests need to be done to study the long-range extrapolation behavior of such models", indicating the rising interest in hybrid neural network modeling. The introduction of physics-informed neural networks (PINNs) \citep{raissi2019} was a significant step to solve forward and inverse problems by involving nonlinear partial differential equations. Since then, several variations and extensions have been proposed such as the physics-informed PointNet (PIPN) \citep{kashefi2022} and extended physics-informed neural networks (XPINNS) \citep{jagtap2020}. \citep{cuomo2022} provides a review on applications and variations of PINNs.

\citet{linka2021} also support data-driven constitutive modeling by continuum-mechanics knowledge in a unified network approach referred to as constitutive artificical neural networks (CANNs). Information on the physical problem helps to reduce the required data set as was also demonstrated by \citet{tang2019}. The data-driven approach suggested by \citet{kirchdoerfer2016} and the application to parameter identification in \citep{schowtjak2022} show a stronger orientation towards the material behavior, i.e., directly to experimental material data and pertinent constraints with conservation laws. A clear motivation for improvements of data-driven approaches can generally be seen in the traditional ideas of model order reduction, e.g., by training of scale separation \citep{unger2009} or reduced homogenization, e.g., for transient diffusion problems \citep{waseem2021}. \citet{settgast2020} demonstrate the usefulness of a hybrid multi-scale neural network approach for the description of yield functions and evolution equations. This also applies to uncertain data and reliability analysis, for instance, during the design stage of artificial neural networks for real-time predictions of mechanized tunneling processes \citep{freitag2018}. 
\clearpage

\subsection{The more physical information, the better the prediction?}
While utilizing such physical information can yield an improvement, this is not granted, though. Even worse, an unfortunate implementation of physically sound constraints into an ANN or other frameworks may lead to poorer predictions. Before a strategy is derived on the basis of Rao-Blackwellization, this should be introduced clearer by a simplified example. The final example of this study will later also discuss the now evident limitations for the practical use of digital image correlation (DIC) data with ANNs in numerical simulations.

\begin{figure}[htp]
\centering
\includegraphics[width=0.75\textwidth]{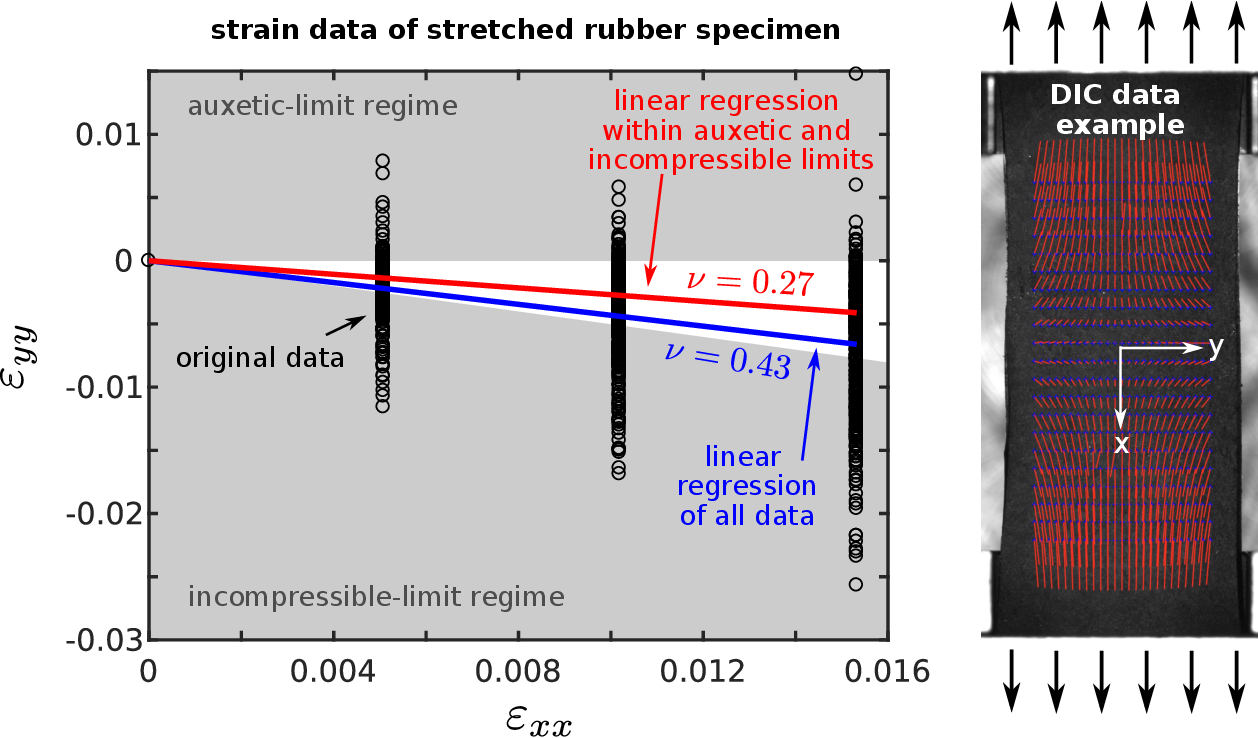}
\caption{Uniaxial stretch of a bike-wheel rubber sample: Poisson's ratio approximation based on original data vs.~data not exceeding incompressible and auxetic limits. An example of DIC-calculated displacements is given by red lines superposed on the sample on the right. Longitudinal strain ($\varepsilon_{xx}$) is determined from clamp displacement, lateral strain ($\varepsilon_{yy}$) is derived from DIC data over multiple regions.}
\label{fig:introductory_example_poisson_ratio}
\end{figure}

Let us assume a nearly-incompressible polymer with Poisson's ratio $\nu$ below but close to 0.5. Experimental data of a stretched bicycle tube patch is shown in Fig.~\ref{fig:introductory_example_poisson_ratio}. Longitudinal strain ($\varepsilon_{xx}$) is determined from reliable displacement measurements at the clamped ends. The limited image resolution and edge detection, however, causes low sensitivity for lateral strain. Lateral strains ($\varepsilon_{yy}$) are thus obtained via DIC at multiple regions of the image. As a result, the $\varepsilon_{yy}$ component thus contains multiple yet noisy points for each stretched state, e.g., approximately of the format $\varepsilon_{yy} = - \nu \varepsilon_{xx} + \epsilon$ with a normal-distribution error $\epsilon \sim \mathcal{N}(0,\sigma^2)$. Some of this data may even exceed the physically sound limits of incompressible and auxetic materials ($\nu = 0.5$ and $\nu = 0$, shaded regions in Fig.~\ref{fig:introductory_example_poisson_ratio}). In a first attempt, only the physically sound data base may be treated as sufficient information. Disregarding data that exceeds the incompressible and auxetic limits may seem a reasonable pre-processing step for the individual points.

Removing this unphysical data, however, worsens the prediction for a subsequent linear regression. The Poisson's ratio is $\nu = 0.23$ for the reduced set of physically admissible data points. Accounting for all data, also the unphysical one, yields a more realistic prediction of $\nu = 0.43$. This is closer to the typical value of soft polymers. Obviously, simply imposing physically sound side constraints does not guarantee improvement. The accessibility of this introductory example fortunately allows to understand the obvious misuse of physical information. Applying the physical constraints to the noise $\epsilon$ of the original data adds a bias, which affects later ANN training or fitting. 

More complex physical systems described by ANNs, however, often do not allow such a clear interpretation. Sources of uncertainties are plentiful and so are the possibilities to implement physical information into ANNs, e.g., into their training data, their structure or their hyperparameters. To apply physical knowledge may be well justified. But asking
\begin{itemize}
\item what information to incorporate?
\item when to incorporate this information?
\item how to incorporate this information?
\end{itemize}
are questions that are not easily assessed. They motivate the present study. It will be shown how intuitive assumptions such as isotropy can help to improve the generation of ANNs, but also where limitations may arise if superfluous or biased information is transferred.

\subsection{Rao-Blackwellization of physical models: sufficient information for guaranteed improvement}

In addition to significant developments in physics-based ANNs, strategies of efficiently employing information have also been elaborated in the area of statistical modeling. The specific inspiration for the present work is the so-called Rao-Blackwell(-Kolmogorov) theorem \citep{blackwell1947, lehmann1998, rao1973, bickel2006}. It relies on knowledge about sufficient information. More precisely, it states that the conditional expectation of an initial estimator, given a sufficient statistic, is a better estimator. It can be proven mathematically that the new estimator, which employs sufficient information, is never worse. This optimality condition and the construction scheme for the new estimator will help to answer the questions raised above. 

Note that statistical tools work on statistical models based on random distributions, though. They are not designed for physical systems and their models. The present work hence combines the mathematical structure of Rao-Black\-wellization with sufficient physics information for the enrichment of artificial neural networks. It is important to note that the proposed, new combination will harness information about the physical model \citep{kurzeja2016} instead of statistical models \citep{doucet2000}. Referring to the term Rao-Blackwellization should therefore not be misleading towards the original concept, but should acknowledge the original idea.
The focus will be on material models in the framework of solid continuum mechanics. By doing so, the presented approach complements existing approaches of physics-based ANNs in three different ways. It
\begin{itemize}
\item does not ask for more information but just for sufficient information,
\item mathematically introduces a rigorous error measure and
\item can be applied even to crude initial predictions.
\end{itemize}
This is particularly beneficial for scenarios with incomplete data sets and where the formats of the solution is secondary, i.e., data-based solutions such as ANNs used in practice.

The present work will first establish a formal framework for Rao-Blackwellization of physical models in section 2. It includes an introductory example based on the micro-sphere model. The general concepts for artificial neural networks are then highlighted by four examples. Section 3 provides an analytical example for predicting a yield function utilizing a description via invariants. Section 4 demonstrates how the database of drilled steel bars undergoing material softening can be sufficiently extended by dimensionless analysis without the need for new measurements. A damage model based on a variational minimization problem allows to compare different realizations of the improvement strategy in section 5 in terms of modifications of ANN output, data and structure. Finally, the numerical adoption in section 6 utilizes isotropic elasticity for the simulation of rubber samples based on DIC and force-displacement data. All examples highlight different aspects of how sufficient information can improve ANNs from data generation, to the ANN structure and its output.

\section{Transferring Rao-Blackwellization from statistical models to deterministic material models}

\subsection{Adaption of the Rao-Blackwellization framework to material models}

Statistical modeling and physical material modeling both share the main challenge of predicting unknown quantities. They are also similar as they both follow the strategy of employing a governing set of mathematical equations supported by modeling assumptions and assessed by error norms. The specific implementations and nomenclature yet differ, which is why we first explain the key differences that are relevant to adapt the Rao-Blackwell theorem. 

While a statistical model builds upon a sample space (e.g.~tensile specimen copies), the material model builds upon physical, deterministic configurations (e.g.~stress states or microstructures). Statistical model parameters can be mean and standard deviation of sustained force, whereas material model parameters can be Young's modulus and Poisson's ratio. On the one hand, a statistical estimator predicts an estimate of a given quantity from observed data (e.g.~a sample mean force). On the other hand, an analogy in deterministic material modeling may be the prediction of an unknown Poisson's ratio but also predictions of stresses or strains work within this framework. Finally, a possible error measure of a statistical estimator can be defined by the expected value $\mathbb{E}$ of the mean squared error. A similar mean squared error can be introduced in material modeling by performing an average $\mathbb{A}$ of the mean squared error over a set of physical configurations. The specific, necessary mathematical definitions and implementation are derived in the following and the interested reader is referred to \citep{kurzeja2016} for more details and a closer look on multi-scale problems.

To become more specific in terms of material modeling, we start from a prediction of an unknown quantity $\theta$ that depends on the physical state $\omega \in \Omega$. For instance, we may want to distinguish elastic loading ($\theta < 0$) from plastic loading ($\theta = 0$) and from inadmissible stress states ($\theta > 0$) for a given stress tensor ($\omega = \boldsymbol\sigma$). Moreover, we assume from a physical constraint that there is sufficient information $\mathcal{S}$ for the prediction of the unknown. For instance, we can assume that the effective macroscopic plastic yielding of polycrystals can often be approximated on the basis of the deviatoric stress norm ($\mathcal{S} = \| \boldsymbol\sigma^{\text{dev}} \|$). Note that we do not need to know how the exact relationship $\theta(\omega)$ looks like. The aim is to find a better one, which is achieved in three steps:
\begin{itemize}
\item[1.] Let $\theta_0(\omega)$ be any starting prediction. 
\item[2.] Let $\mathcal{S}$ be sufficient information to determine $\theta$. 
\item[3.] Then, the new prediction
\begin{align}
\theta_1(s) = \mathbb{A}\left(\theta_0 \, \vert \, \mathcal{S} = s\right) = \frac{1}{\vert \Omega_s \vert} \int\limits_{\Omega_s} \theta_0 \, \diff \omega, \label{eq:theta1}
\end{align}
is never worse in the sense of the averaged mean squared error
\begin{align}
\frac{1}{\vert \Omega \vert} \int\limits_\Omega (\theta_1 - \theta)^2 \, \diff \omega
\le
\frac{1}{\vert \Omega \vert} \int\limits_\Omega (\theta_0 - \theta)^2 \, \diff \omega. \label{eq:error_inequality}
\end{align}
Therein, $\Omega_s:=\{ \omega \in \Omega \, | \, \mathcal{S}(\omega) = s\}$ is the subdomain, where $\mathcal{S}(\omega)$ takes the constant value $s$. $\vert \Omega \vert: = \int_{\Omega} 1 \, \diff \omega$ and $\vert \Omega_s \vert: = \int_{\Omega_s} 1 \, \diff \omega$ are used for normalization.
\end{itemize}
In brief, $\theta_1$ may be interpreted as a conditional averaging of the initial prediction $\theta_0$ over constant values $s$ of the sufficient information $\mathcal{S}$. This interpretation is different from the statistical use of an expected value. What is most important, the physical constraint tells us that the prediction of $\theta$ must be fully determined by prescribing $s$. For instance, two stress states $\boldsymbol\sigma_I$ and $\boldsymbol\sigma_{II}$ must be both elastic or both plastic if they have the same deviatoric stress norm, $ \| \boldsymbol\sigma_I^{\text{dev}} \| = s_I = s_{II} = \| \boldsymbol\sigma_{II}^{\text{dev}} \| $. The adopted Rao-Blackwell theorem thus averages over such states for the new prediction. As a first key result, we did not only obtain a strategy to incorporate physical knowledge --- it is guaranteed not to worsen our initial prediction. A fundamental property of this scheme is that the error does not grow. The proof is sketched in appendix A.

\subsection{Notes on the practical implementation of sufficient information}
The strategy employs three steps --- initial estimator, sufficient set, improved estimator --- each with individual practical peculiarities. Various initial predictions can be tested in application, e.g., neural networks trained with different initial weights, data or transfer functions. The initial prediction can even be a crude one, while two special cases would trivially not yield an improvement in the sense of a strict inequality \eqref{eq:error_inequality}; namely, the correct prediction (it cannot be improved further) and an already improved prediction (due to idempotency).

For the second step of the scheme, identifying a sufficient set is just the necessary information for the improvement scheme to properly work. While sufficiency in statistics is proven based on the statistical model properties (c.f.~the Fisher-Neyman factorization \citep{fisher1922,bickel2006}), sufficiency for the material model's predictions is based on knowledge about the physical behavior. Note that sufficiency does not imply sufficient information in every detail. Instead and like the statistical motivation, sufficiency is understood with respect to a model. Obvious sources determining a sufficient set in material modeling can be assumptions like isotropy, linearity or hyperelasticity. Conservation laws, observer invariance or the second law of thermodynamics can also reveal whether a parameter is allowed to have an impact on another. Other sources of sufficiency can be the parameters of a set of governing equations, which sufficiently fix the unknown solution. A dimensional or homogenization analysis can moreover help to reduce the set of required parameters for prediction. This list is not exhaustive and shall provide a few ideas for practical application.

Having sufficient information at hand, equation \eqref{eq:theta1} defines the third step, i.e., the improved prediction. Three possibilities of its employment into the ANN are, for instance,
\begin{itemize}
\item ANN training data augmentation,
\item adapting the ANN structure,
\item filtering the ANN output
\end{itemize}
and will be shown in the following examples.
In the ideal case, the integral in \eqref{eq:theta1} is evaluated analytically but may be unattainable in practice. In such cases, a numerical approximation of the scheme is required. The idea is to transform the continuous parameterization of $s$ into discrete intervals. The discretization highly depends on the range of $s$, which may be unbounded or bounded by physical constraints. An intuitive approach for the latter case is to introduce $n + 1$ intervals
\begin{align}
s_i = \left[ s_{\mathrm{min}} + \frac{i}{n+1} \, \left( s_{\mathrm{max}} - s_{\mathrm{min}} \right) , s_{\mathrm{min}} + \frac{i+1}{n+1} \, \left( s_{\mathrm{max}} - s_{\mathrm{min}} \right) \right),
\end{align}
with $i = 0, \dots ,n$. With these at hand, the integral can be replaced by a sum over all data points $m$, which fall into the interval $s_i$, to determine the new model
\begin{align}
\theta_1(s_i) = \frac{1}{m} \sum\limits_{j = 1}^{m} \theta_0\left(\mathcal{S}\left(\omega\right) \in s_i \right).
\end{align}
As one result, the numerical approximation of the improved prediction may not fully inherit the optimality condition. It remains a compromise between interval stepping and averaged data, limited by data availability and computational capacities.  


\subsection{Introductory example for Rao-Blackwellization of physical constraints: the micro-sphere model} \label{subsec:microsphere}

An example from the area of micro-sphere modeling \citep{miehe2004} shall make the concept of Rao-Blackwellization with physical constraints comprehensible. It is purely analytical and not combined with artificial neural networks or numerical evaluations yet. Statistical considerations of micro-sphere modeling (such as Langevin statistics and random-walk of chain segments) are also omitted for better distinction of physics-based Rao-Blackwellization from statistics-based Rao-Blackwellization. Like all examples presented herein, the introductory example follows the same three steps: 
\begin{enumerate}
\item initial estimator $\theta_0$,
\item sufficient information $\mathcal{S}$,
\item improved estimator $\theta_1$.
\end{enumerate}
For the sake of simplicity, the symbols $\theta_0$, $\mathcal{S}$ and $\theta_1$ will be used in all examples without further modification of the notation.

\begin{figure}[h]
\centering
\includegraphics[width=0.9\textwidth,fbox]{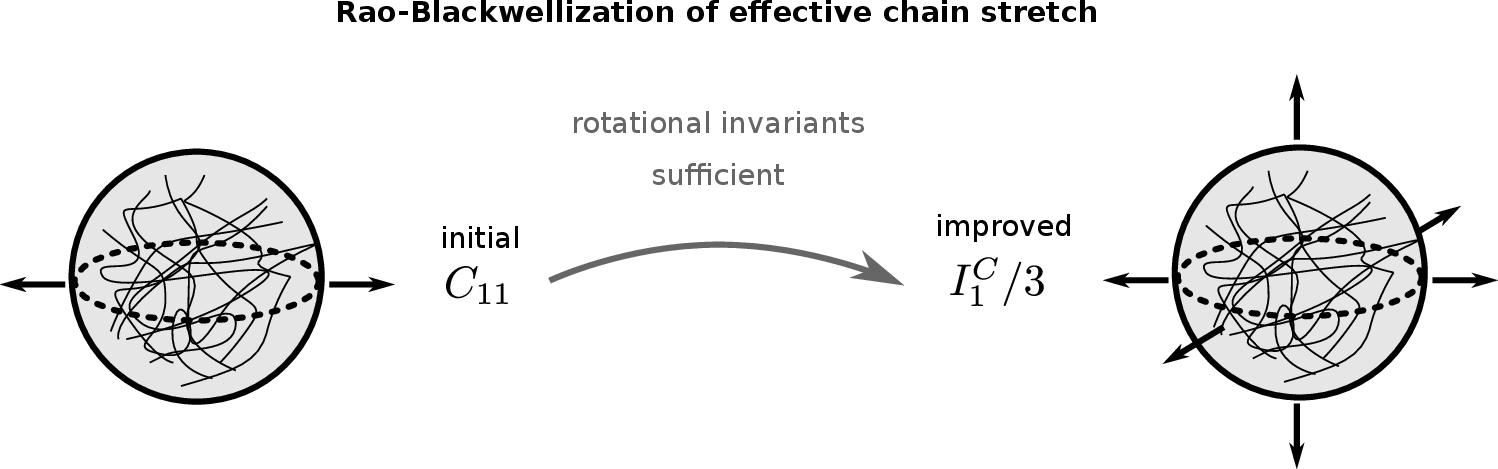}
\caption{Example of Rao-Blackwellization within the chain-based micro-sphere framework: chain stretch from one in three directions. Improvement is based on sufficient information under the assumption of rotational invariance.}
\label{fig:introductory_example}
\end{figure}

Consider a rubber network of polymer chains that undergoes macroscopic deformation described by the right Cauchy-Green tensor $\textbf{C}$, Fig.~\ref{fig:introductory_example}. The task is to predict an effective, quadratic stretch of the underlying polymer chains $\lambda_c^2$. This effective stretch is commonly required for the energy functions of rubber models \citep{miehe2004,arrudaboyce1993}. We now have to pick some information from $\textbf{C}$ to predict $\lambda_c^2$. 

While there are a few obvious ad-hoc choices, some even considering superposed microstructural fluctuations, we deliberately start with a poor initial estimator. The initial estimator predicts the effective quadratic stretch of the polymer chains by the 11-component only of some cartesian coordinate system
\begin{equation}
\theta_0 = C_{11} = \textbf{e}_1 \cdot \textbf{C} \cdot \textbf{e}_1.
\end{equation}
This initial estimator is the first step of Rao-Blackwellization but only acceptable for uni-axial stretch. It clearly fails for biaxial tension, for instance. We would therefore like to reduce the error for all possible deformation states.

We now further assume that the assumption of isotropy is a valid physical approximation, e.g, based on fabrication conditions and experimental indication. The second step of Rao-Blackwellization then requires the sufficient information under this constraint. For an isotropic model, it is sufficient to know the three (rotational) invariants of the strain tensor $\textbf{C}$. These invariants thus constitute the sufficient set
\begin{equation}
\mathcal{S} = \{I^C_1, I^C_2, I^C_3 \}.
\end{equation}

The third step of Rao-Blackwellization now yields the improved prediction by averaging over the domain of deformation states, where the sufficient set is constant. This subdomain of deformation states, where $\mathcal{S} = \{I^C_1, I^C_2, I^C_3 \}$ is constant, is given by just the rotations $\textbf{Q}$ of the strain tensor:
\begin{align}
 \Omega_s :&= \{ \text{all }\textbf{C} \text{ with same values of invariants } I^C_1, I^C_2, I^C_3 \} \nonumber \\
 &= \{ \textbf{Q}^T \cdot \textbf{C} \cdot \textbf{Q}: \, \forall \textbf{Q} \in \text{SO(3)} \text{ and starting } \textbf{C} \text{ with invariants } I^C_1, I^C_2, I^C_3 \}. \nonumber 
\end{align}
 
The improved estimator hence reads (see \citep{miehe2004} for the algebra of the last step)
\begin{align}
\theta_1 (I^C_1, I^C_2, I^C_3) \nonumber 
&=
\frac{
\int\limits_{\Omega_s (I^C_1, I^C_2, I^C_3)} \theta_0 (C_{11})  \, \mathrm{d}\omega
}{
\int\limits_{\Omega_s (I^C_1, I^C_2, I^C_3)} 1  \, \mathrm{d}\omega
}
=
\frac{
\int\limits_{\Omega_s (I^C_1, I^C_2, I^C_3)} \textbf{e}_1 \cdot \textbf{C} \cdot \textbf{e}_1  \, \mathrm{d}\omega
}{
\int\limits_{\Omega_s (I^C_1, I^C_2, I^C_3)} 1  \, \mathrm{d}\omega
} \nonumber \\
&=
\frac{
\int\limits_{0}^{2 \pi} \int\limits_{0}^{\pi} \textbf{e}_1 \cdot \textbf{Q}^T(\vartheta,\varphi) \cdot \textbf{C} \cdot \textbf{Q}(\vartheta,\varphi) \cdot \textbf{e}_1 \, \sin \vartheta \, \mathrm{d}\vartheta \, \mathrm{d}\varphi
}{
\int\limits_{0}^{2 \pi} \int\limits_{0}^{\pi} 1 \, \sin \vartheta \, \mathrm{d}\vartheta \, \mathrm{d}\varphi
} \nonumber \\
&=
\frac{
\int\limits_{0}^{2 \pi} \int\limits_{0}^{\pi} (\textbf{Q}(\vartheta,\varphi) \cdot \textbf{e}_1) \cdot \textbf{C} \cdot (\textbf{Q}(\vartheta,\varphi) \cdot \textbf{e}_1) \, \sin \vartheta \, \mathrm{d}\vartheta \, \mathrm{d}\varphi 
}{4 \pi}
\nonumber \\
&=
\frac{C_{11} + C_{22} + C_{33}}{3} = \frac{I^C_1}{3}. 
\end{align}
The improved estimator thus simplifies to a third of the first invariant $\theta_1 = I^C_1$. This is just the wanted effective, quadratic stretch $\lambda_c^2$ for the limit case of the eight-chain Arruda-Boyce model when derived from the micro-sphere framework \citep{miehe2004,arrudaboyce1993}. Application to other powers of the stretch, e.g.~a linear estimation, may yield to different weighting factors and we refer to \citep{miehe2004} for details on the micro-sphere approach and to \citep{kurzeja2018} for another analyical example of physics-based Rao-Blackwellization. 

Rao-Blackwellization however stands out from a micro-sphere approach by some significant characteristics and another scope of application. Instead of developing an entire framework, it just employs an initial estimator (even an improper one is possible) and knowledge about sufficient information on the unknowns (which is the point where the additional physics enter the prediction). This makes it a suitable tool for situations where a complete physical framework is not provided, e.g., predictions based on artificial neural networks as demonstrated by the following scenarios.

\section{Illustrating Rao-Blackwellization effects on an ANN - Example of an elasto-plastic predictor step} \label{sec:yieldfunction}

To introduce the application of the proposed strategy to an ANN, the yield function defining the transition between a fully elastic and an elasto-plastic state is to be identified. Accordingly, the ANN shall predict whether a deformation will fall into the elastic regime or not. Point of departure is the classic von Mises yield criterion in the 2d stress space
\begin{align}
\Phi^{\text{vM}}(\sigma_1, \sigma_2) = \sqrt{\sigma_1^2 + \sigma_2^2 - \sigma_1 \, \sigma_2} - \sigma_Y 
\end{align}
that is formulated in the principal stresses $\sigma_1$ and $\sigma_2$ and can be motivated for shear-dominated plasticity in crystals. The analytical starting point will also serve as a reference solution and eases an interpretation of the later improvement. The material's yield stress is chosen to be $\sigma_Y = 1 \, \text{MPa}$. Considering an elastic predictor step within a numerical implementation (e.g., the return-mapping scheme), the different states can be identified as 
\begin{align}
	\Phi^{\text{vM}}(\sigma_1, \sigma_2) 
	\quad \rightarrow
	\begin{cases}
		\le 0 \text{ (elastic)}\\
		> 0 \text{ (inadmissible $\Rightarrow$ elasto-plastic corrector).}
		\label{eq:ex1-vonmises}
	\end{cases}
\end{align}

\subsection{Initial ANNs distinguishing elasticity from plasticity}

The initial ANN,  $\theta_0$, shall predict whether a given stress pair $(\sigma_1,\sigma_2)$ corresponds to a purely elastic step or not. We test three initial ANNs of different sizes, which are layer-wise: 5-1, 10-5-1 and 200-200-60-1. All neurons are fully connected to the neighboring layers and activated by the hyperbolic tangent (tanh) function. The output of the last layer is moreover processed such that the discrete distinction between elasticity and plasticity is numerically labeled by 0 and 1 in the form (cf. Eq.~\eqref{eq:ex1-vonmises})
\begin{align}
o = \begin{cases}
0 & \text{if} \, \,  \theta_0(\sigma_1,\sigma_2) \le 0 \quad \text{ (elastic) } \\
1 & \text{if} \, \,  \theta_0(\sigma_1,\sigma_2) > 0 \quad \text{ (inadmissible $\Rightarrow$ elasto-plastic corrector).} 
\end{cases}
\end{align}

\begin{figure}[ht] 	
	\psfrag{A}[c][c]{stress $\sigma_1$ [MPa]}
	\psfrag{B}[c][c]{stress $\sigma_2$ [MPa]}
	\psfrag{C}[l][c]{elastic training data}
	\psfrag{D}[l][c]{plastic training data}
	\psfrag{E}[l][c]{elastic validation data}
	\psfrag{F}[l][c]{plastic validation data}
	\psfrag{1}[c][c]{1}
	\psfrag{0}[c][c]{0}
	\psfrag{-1}[c][c]{-1}
	\centering
     \includegraphics[width=0.5\textwidth]{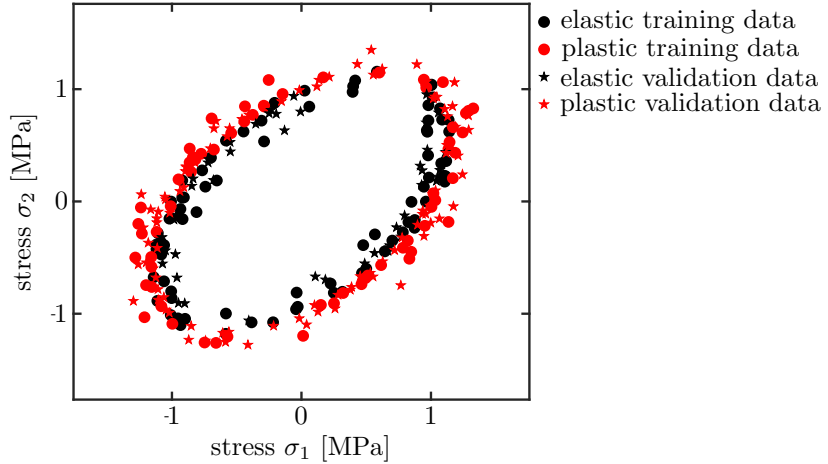}
  \caption{Training and validation data for the ANNs distinguishing elastic and plastic predictor steps.}
  \label{fig:ex1-trainingdata}
\end{figure}

Data was generated from the analytical solution in the range of $\Omega = [-1.75\,\text{MPa},+1.75\,\text{MPa}]^2$ for training, validation and testing, see Fig.~\ref{fig:ex1-trainingdata}. Training and validation data was moreover superposed with random noise to highlight the influence of imperfect data. Stress states that fall into the range $[-0.03\,\text{MPa},+0.03\,\text{MPa}]$ of the yield function were assigned randomly. For the test data, the stress domain $\Omega$ was discretized by $\Delta \sigma = 0.01 \, \text{MPa}$. No regularization of the input data is required in this example by working with units of MPa. The networks have been finally trained via the Stochastic Gradient Descent (SGD) method and the network performance was evaluated by the mean squared error. A detailed summary of the ANN hyperparameters and settings is provided in Appendix \ref{chp_appendix} for all examples.

The initial ANN of size 5-1 shows a significant error due to its simplicity that causes underfitting. Unphysical predictions can be observed in the form of separated and non-convex elastic regimes, see the initial (grey) performance in Fig.~\ref{fig:ex1-err-rb}. The 10-5-1 network performs much better, matching both the qualitative and the quantitative prediction of the elastic-plastic regimes. The 200-200-60-1 network achieves an even lower error in the training set, but the validation and test set errors are higher compared to the 10-5-1 case. This overfitting behavior is caused by the high number of degrees of freedom. Given the initial prediction of all networks ($\theta_0$), all three shall be optimized via physics-based Rao-Blackwellization. 

\subsection{Sufficient physical information for the yield condition}
Motivated by slip in crystals, for instance, we assume that the onset of plasticity is dictated by the deviatoric stress norm. For the present principal stresses and analytical reference, it writes
\begin{equation}
\mathcal{S} (\sigma_1,\sigma_2)
:=
\| \boldsymbol \sigma^{\text{dev}} \|
 = 
\sqrt{\sigma_1^2 + \sigma_2^2 - \sigma_1 \, \sigma_2}
\end{equation}
and completely determines whether the material undergoes elastic or plastic deformation. It is thus \textit{sufficient} to distinguish between both cases under this assumption. The simple, underlying analytical solution in \eqref{eq:ex1-vonmises} is provided here as an idealized benchmark to be met for better interpretation of the results. Unpredictable solutions follow in the later examples. Also note at this point that we do not need to know how this sufficient information relates to the prediction. The Rao-Blackwell scheme only receives an imperfect ANN as an initial predictor and the sufficient set of information.

\subsection{Improved ANNs distinguishing elasticity from plasticity}

The improved ANNs, $\theta_1$, are calculated as, cf.~\eqref{eq:theta1},
\begin{align}
&\theta_1(\sigma^{\text{dev}}) = \frac{1}{\vert \Omega_{\sigma^{\text{dev}}} \vert} \int\limits_{\Omega_{\sigma^{\text{dev}}}} \theta_0 \, \diff \sigma ,\\
&\Omega_{\sigma^{\text{dev}}} = \left\{ \{\sigma_1,\sigma_2\} \in [-1.5\,\text{MPa},+1.5\,\text{MPa}]^2 \, \text{ with } \, \| \boldsymbol \sigma^{\text{dev}} (\sigma_1,\sigma_2) \| = \sigma^{\text{dev}} \right\} .
\end{align}
They hence only depend on a single scalar value $\sigma^{\text{dev}}$. They derive from the initial ANNs (depending on $(\sigma_1,\sigma_2)$) by conditional averaging over all stress states with the same effective stress value $\sigma^{\text{dev}}$, which are denoted by the subdomain $\Omega_{\sigma^{\text{dev}}}$. Potential reduction of the input parameters is already an important outcome for the application to neural networks. 

The initial ANNs provide discrete distributed data but no solution that allows for an explicit analytical integration. We thus approximate the stress domain of the improvement scheme numerically by a discretization with intervals
$
\sigma^{\text{dev}}_i = [ \left( i - 0.5 \right) \, ds, \left( i + 0.5 \right) \, ds ],
$
 $i = 0, \dots ,n$. The integral is then replaced by a sum over all $m$ data points that fall into the range of $\sigma^{\text{dev}}_i$. Interpolation is used for empty intervals and the interval number is limited to $n = 1750$. Just like for the initial ANNs, the output of the new ANNs is also rounded to 0 or 1 to obtain a clear distinction between elastic and plastic predictors, respectively.

\begin{figure}[ht] 
	\centering
	\psfrag{A}{initial ANN}
	\psfrag{B}{improved ANN}
	\psfrag{C}[c][c]{mean squared error}
	\psfrag{D}[c][c]{5-1}
	\psfrag{E}[c][c]{10-5-1}
	\psfrag{F}[c][c]{200-200-60-1}
	\psfrag{1}[c][c]{$10^{-4}$}
	\psfrag{2}[c][c]{$10^{-3}$}
	\psfrag{3}[c][c]{$10^{-2}$}
	\psfrag{4}[c][c]{$10^{-1}$}
     \includegraphics[width=0.82\textwidth]{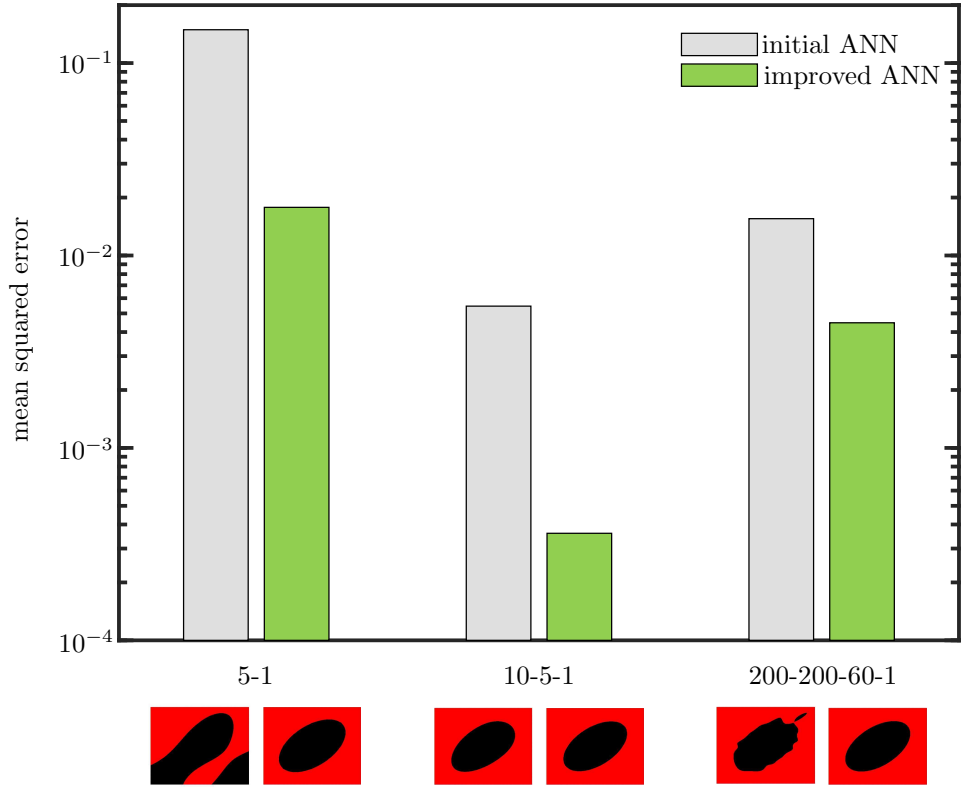}
  \caption{Mean squared errors determined by the test set and yield functions for different network sizes before and after Rao-Blackwellization.}
  \label{fig:ex1-err-rb}
\end{figure}

Physics-based Rao-Blackwellization improves all initial networks' output (sizes 5-1, 10-5-1 and 200-200-60-1) significantly in terms of the cost function and the shape of the flow curves, Fig.~\ref{fig:ex1-err-rb}. Underfitting of the smallest network initially caused completely unphysical predictions that are now well solved by a single, convexified elastic domain. The error dropped by a factor of eight. Also overfitting of noise in the largest network is now successfully suppressed by an error improvement of more than factor 3. Even the very good prediction of the middle-sized network was further improved notably by a factor of 15 in terms of the cost function. The physics-based Rao-Blackwell strategy hence shows clear improvements of all ANNs. Smoothing of noise and overfitting are two illustrative outcomes in this first example. 

\section{Splitting computational costs by dimensional analysis: Ultimate tensile force in a drilled steel bar}

This example aims at adapting the implementation of the ANN within the overall framework of an engineering problem. Some physical knowledge indeed suggests a reformulation of the overall problem instead of improving single parts such as the input layer or the output. This will be demonstrated by a dimensional analysis of a steel bar with a drill hole. To be more precise, the role of the ANN will be reduced to solve the dimensionless formulation of the problem. The dimensional scaling is outsourced to an external calculation. Note that a dimensionless treatment is a well-known and successful concept, as numerical simulations and ANNs inherently perform dimensionless computations. Yet several industrial and academic projects work with dimensional descriptions due to methodological or practical reasons. The following physics-based Rao-Blackwellization will show the potential of significant cost reduction by a dimensional analysis in terms of data and computational resources. In addition to the well-established dimensional analysis, the example will demonstrate how even coarse data sets allow substantial improvement. Data is obtained from elasto-plastic finite-element simulations, which makes data generation costly.

\begin{figure}[h]
\centering
	\psfrag{A}[c][c]{displacement $u$ [mm]}
	\psfrag{B}[c][c]{force $F$ [kN]}
	\psfrag{F1}[c][c]{$F_{\text{y}}$}
	\psfrag{F2}[c][c]{$F_{\text{uts}}$}
	\psfrag{W}[c][c]{$W$}
	\psfrag{w}[c][c]{$w$}
	\psfrag{R}[c][c]{$R$}
	\psfrag{d}[c][c]{$d$}
	\psfrag{L1}[c][c]{$L_\mathrm{i}$}
	\psfrag{L2}[c][c]{$L_\mathrm{o}$}
	\psfrag{0}[c][c]{0}
	\psfrag{0.2}[c][c]{0.2}
	\psfrag{0.6}[c][c]{0.6}
	\psfrag{0.4}[c][c]{0.4}
	\psfrag{0.8}[c][c]{0.8}
	\psfrag{2}[c][c]{2}
	\psfrag{4}[c][c]{4}
	\psfrag{6}[c][c]{6}
	\psfrag{8}[c][c]{8}
	\psfrag{10}[c][c]{10}
\includegraphics[width=0.9\textwidth]{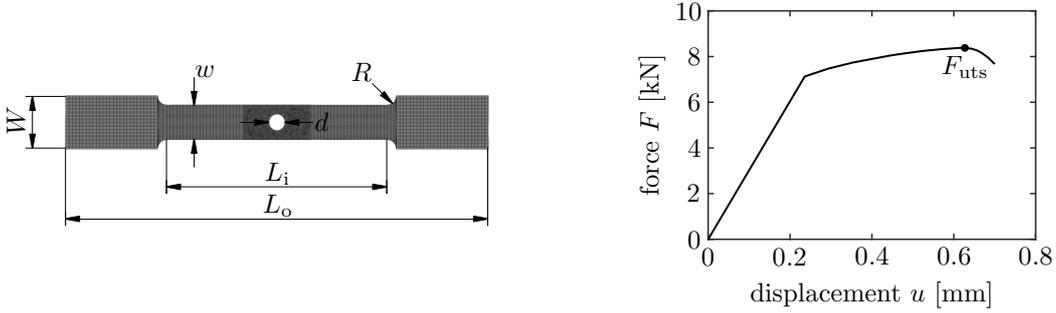}
\caption{Tension of a drilled steelbar: sketch of testing specimen (left) and characteristic force of ultimate tensile strength to be predicted for an exemplary specimen width of $w = 4$ mm and drill diameter of $d = 2$ mm (right).}
\label{fig:drilled_steelbar_problem}
\end{figure}

The bar in this example shall undergo elasto-plastic deformation and the material is assumed to be a high-strength dual phase steel. Due to different application scenarios, the geometry varies with respect to width of the square cross section (2 mm $< w <$ 8 mm) and drill hole diameter ($d < w$). We are interested in the tensile force at the point of ultimate tensile strength ($F_\text{uts}$) for the entire range of products, i.e., for all possible combinations of width $w$ and drill hole diameter $d$, see Fig.~\ref{fig:drilled_steelbar_problem}.
\subsection{Initial prediction from simulations}
The initial prediction in this example is a data base calculated from a series of simulations using Abaqus Implicit. They provide force values for each geometric combination of width and drill hole diameter, $w$ and $d$. The specimens are stretched in horizontal direction via Dirichlet boundary conditions on both sides and a Gurson-Tvergaard-Needleman (GTN) damage model combined with Swift hardening is implemented to model the material behavior. The material hence undergoes hardening followed by the onset of degradation. While a regularization method is required to capture localized softening \citep{simo1993,peerlings2001,langenfeld_2022}, it is not needed for the present determination of the ultimate tensile strength. The model parameters are chosen according to \citep{schowtjak2020} for a dual phase steel. All parameters determining the problem are summarized in Tab.~\ref{tab:dimensionless_parameters} in dimensional and dimensionless format.

The time-consuming simulations can however only provide predictions for individual specimens. An ANN may thus serve as a time-efficient, interpolating alternative.
For this matter we investigate three different training data sets of 10 points each, see Fig.~\ref{fig:drilled_steelbar_example_results} for a visualization of all data sets. The first one has a constant drill hole diameter of $d= 4 \,\text{mm}$ with equally spaced, varying width in the range of $\left[4.76,8\right]\, \text{mm}$. The second one is characterized by a constant specimen width of $w = 4 \text{mm}$ and equally spaced drill hole diameters $d$ in the range of $\left[0.364,3.64\right] \,\text{mm}$. The third training data set consists of randomly distributed data points. The test set comprises a finer sampling of the domain with 21 data points. 

\begin{figure}[h]
\begin{center}
\fbox{%
\begin{tikzpicture}[node distance=3cm]
\tikzstyle{network} = [rectangle, rounded corners, minimum width=2.5cm, minimum height=1cm,text centered, draw=black, fill=blue!5]
\tikzstyle{io} = [trapezium, trapezium left angle=70, trapezium right angle=110, minimum width=2.5cm, minimum height=1cm, text centered, draw=black, fill=blue!30]
\tikzstyle{data} = [rectangle, minimum width=2.5cm, minimum height=1cm, text centered, draw=black, fill=white!30]
\tikzstyle{constdata} = [rectangle, minimum width=2.5cm, minimum height=1cm, text centered, draw=black, fill=black!5]
\tikzstyle{decision} = [diamond, minimum width=2.5cm, minimum height=1cm, text centered, draw=black, fill=green!30]
\tikzstyle{arrow} = [thick,->,>=stealth]
\node (input) [data, align=left] {$w$\\ $d$};
\node (ann) [network, right of=input] {ANN$_\text{dim}$};
\node (output) [data, align=left, right of=ann] {$F_\text{uts}$};
\node (constinput) [constdata, align=center, below of=input, yshift=1.5cm] {material parameters: constant};
\draw [arrow] (input) -- (ann);
\draw [arrow] (ann) -- (output);
\end{tikzpicture}
}
\end{center}
\caption{Structure of the initial, dimensional ANN$_\text{dim}$ for the prediction of the characteristic force.}
\label{fig:ANN_dimensional}
\end{figure}
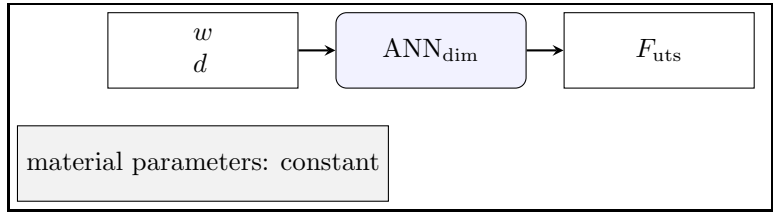

Following the original, dimensional data structure of the simulations, a straight-forward ANN has the form as shown in Fig.~\ref{fig:ANN_dimensional}. This ANN$_\text{dim}$ varies the same input parameters like the original simulations, $w$ and $d$. While providing a quick approximation and interpolation of the simulation data, it neglects the sufficient information from a dimensionless description. We will thus compare it to an improved ANN after identifying sufficient information for the dimensionless problem. All ANNs in this example share the same network structure (13-13-13-1), except for an adapted input size, with ReLu activation and are trained via SGD and/or Adam method. For each ANN twenty different initial weight distributions are trained and the one with the best performance is picked. 
\subsection{Sufficient localized and dimensionless information}
Motivated by Rao-Blackwellization to focus only on sufficient information, we try to reduce the parameter set. It is obvious (but should still be mentioned) that constant parameters such as Poisson's ratio are not used as input for the ANN, because the ANN inherently adapts to these constant background conditions during the training process. A first physics-based reduction of input parameters is possible due to localization triggered by the onset of material softening. The prescribed geometric lengths ($L_\text{o}$, $L_\text{i}$, $W$ and $R$) are hence assumed to not interfere with the results as long as $L_\text{i}$ is larger than the zone of plasticity and material softening around the drill hole. The associated lengths and their dimensionless numbers can thus be ignored in the following. As a first result, there must be a relationship between width $w$, diameter $d$ and the force of the form
\begin{equation}
f_\text{dim} \, (w, \, d, \, \, F_\text{uts}) = 0.
\label{eq:dimensional_relationships}
\end{equation}
Note that we do not need to know how this relationship looks like for physics-based Rao-Black\-wellization. It is just important to know that there is a relationship between these parameters such that $w$ and $d$ fully determine the characteristic force (and assuming that the material parameters are out of the discussion, because they are implicitly prescribed and constant). 

A second physics-based optimization of the ANN structure can be achieved from a dimensional analysis. With two units involved in the present system (length and force), the Buckingham $\Pi$-theorem reduces the set of parameters by two. The present setting uses Young's modulus $E$ and bar width $w$ for the purpose of dimensional scaling, which is outsourced from the ANN calculations. For the problem at hand, we can infer that there is a relationship of the form
\begin{equation}
f_\text{dim-less} \, (d/w, \, F_\text{uts}^*) = 0,
\label{eq:dimensionless_relationships}
\end{equation}
with the dimensionless force $F_\text{uts}^*$ = $F_\text{uts}/(E \, w^2)$ at the point of ultimate tensile strength. 
In the dimensionless format, we see only one geometric input parameter left, which allows to utilize the simulation results efficiently by a dimensionless ANN.

\begin{table}[hbt]
\begin{tabular}{llll|l}
\hline
dimensional & unit & value & comment & dimensionless\\
parameters & & & & parameters\\
\hline
\hline
\multicolumn{5}{c}{Prescribed material}\\
\hline
$E$ & [N/mm$^2$] & $210000$ & Young's modulus  & -\\
$\nu$ & [-] & $0.3$ & Poisson's ratio & $\nu$ \\
$\varepsilon_0$ & [-] & $4.044 \cdot 10^{-4}$ & Swift strain hardening parameter & $\varepsilon_0$\\
$A$ & [N/mm$^2$] & $1336.6$ & Swift parameter & $A/E$\\
$n$ & [-] & $0.161$ & Swift exponent & n\\
$q_1$, $q_2$, $q_3$ & [-] & 1.5, 1.0, 2.25 & GTN parameters & $q_1$, $q_2$, $q_3$\\
$f_N$ & [-] & $0.075$ & scales nucleated volume fraction & $f_N$\\
$\varepsilon_N^p$ & [-] & 0.1854 & mean plastic strain at max. nucleation & $\varepsilon_N^p$\\
$s_N$ & [-] & $0.2933$ & standard deviation of nucleation & $s_N$\\
$k_w$ & [-] & 1.0 & shear parameter for nucleation & $k_w$\\
\hline
\multicolumn{5}{c}{Prescribed geometry}\\
\hline
$L_\text{o}$ & [mm] & 46 & &$L_\text{o}/w$ \\
$L_\text{i}$ & [mm] & $24$ & &$L_\text{i}/w$ \\
$R$ & [mm] & $1$ & &$R/w$ \\
\hline
\multicolumn{5}{c}{Varied geometry}\\
\hline
$w$ & [mm] & $\in \, ]0,8]$ & &- \\
$d$ & [mm] & $<w$ & &$d/w$ \\
$W$ & [mm] & $w + 2$ & &$W/w$ \\
\hline
\multicolumn{5}{c}{Resulting characteristic forces}\\
\hline
$F_\text{uts}$ & [N] & & & $F_\text{uts}^*$ = $F_\text{uts}/(E \, w^2)$ \\
\hline
\end{tabular}
\label{tab:dimensionless_parameters}
\caption{Dimensional and dimensionless parameters governing the characteristic force of the drilled steel bar. Model parameters according to a Gurson-Tvergaard-Needleman (GTN) damage model combined with Swift hardening according to \citep{schowtjak2020}.}
\end{table}

\clearpage

\subsection{Improved ANN}

\begin{figure}[h]
\begin{center}
\fbox{%
\begin{tikzpicture}[node distance=3cm]
\tikzstyle{network} = [rectangle, rounded corners, minimum width=2.5cm, minimum height=1cm,text centered, draw=black, fill=blue!5]
\tikzstyle{io} = [trapezium, trapezium left angle=70, trapezium right angle=110, minimum width=2.5cm, minimum height=1cm, text centered, draw=black, fill=blue!30]
\tikzstyle{data} = [rectangle, minimum width=2.5cm, minimum height=1cm, text centered, draw=black, fill=white!30]
\tikzstyle{constdata} = [rectangle, minimum width=2.5cm, minimum height=1cm, text centered, draw=black, fill=black!5]
\tikzstyle{decision} = [diamond, minimum width=2.5cm, minimum height=1cm, text centered, draw=black, fill=green!30]
\tikzstyle{arrow} = [thick,->,>=stealth]
\node (input) [data, align=left] {$\displaystyle d/w$};
\node (ann) [network, right of=input] {ANN$_\text{dim-less}$};
\node (outputdimless) [data, align=left, right of=ann] {$F^*_\text{uts}$};
\node (output) [data, align=left, right of=outputdimless] {$F_\text{uts}$};
\node (inputw) [data, align=center, below of=input, yshift=1.5cm] {$w$};
\node (constinput) [constdata, align=center, below of=inputw, yshift=1.5cm] {material parameters: constant};
\draw [arrow] (input) -- (ann);
\draw [arrow] (ann) -- (outputdimless);
\draw [arrow] (inputw) -| (output);
\draw [arrow] (outputdimless) -- (output);
\draw [arrow] (constinput) -| node[near start] [above]{$E$} (output);
\end{tikzpicture}
}
\end{center}
\caption{Structure of the improved, dimensionless ANN$_\text{dim-less}$ for the prediction of the characteristic force.}
\label{fig:ANN-dimless}
\end{figure}
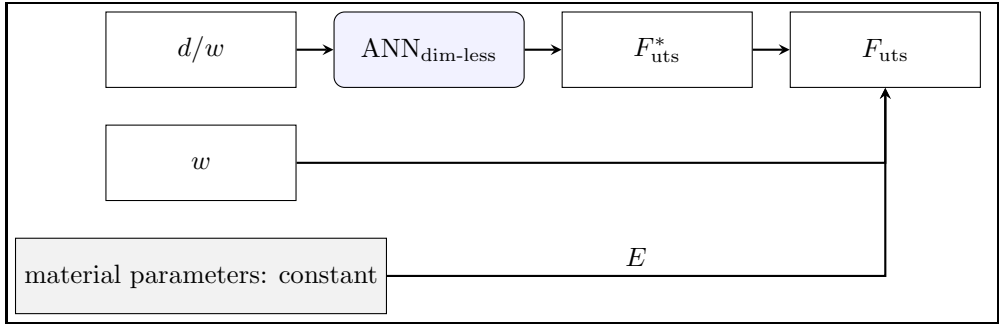

Knowing the dimensionless formulation in \eqref{eq:dimensionless_relationships} allows us to pass the information through the ANN more efficiently. It is denoted as ANN$_\text{dim-less}$ and its structure is illustrated in Fig.~\ref{fig:ANN-dimless}. Sufficient information on the dimensionless force is simply provided by the single value of
\begin{equation}
\mathcal{S} = \frac{d}{w}.
\end{equation}
Rao-Blackwellization of the dimensionless simulation data is thus a conditional average over drill diameters $d$ and widths $w$ of the same ratio
\begin{equation}
F^*_\text{uts} (d/w = s) =
\frac{
\displaystyle \sum_{\text{all pairs }\{w_i, \, d_j\} \text{ with } d_j/w_i = s} \frac{F^{\text{simulation}}_{i,j}}{E \, w_i^2}
}{
\displaystyle \sum_{\text{all pairs }\{w_i, \, d_j\} \text{ with } d_j/w_i = s} 1
}.
\end{equation}

The advantage of the dimensionless ANN is the reduction of complexity and simulation data. The ANN$_\text{dim-less}$ only requires the $d/w$ ratio instead of multiple $w-d$ pairs. The disadvantage of using dimensionless forces is the need for further parameters to obtain the dimensional value. This comes with little costs, though. The dimensional scaling is computationally simple and can indeed be performed outside of the ANN with little effort. The adopted Rao-Blackwell theorem moreover guarantees that the error is never worse in the $w-d$ domain. 

\begin{figure}[h]
\centering
	\psfrag{O}[c][c]{specimen width $w$ [mm]}
	\psfrag{P}[c][c]{drill hole diameter $d$ [mm]}
	\psfrag{R}[l][c]{constant width}
	\psfrag{Q}[l][c]{constant drill diameter}
	\psfrag{S}[l][c]{random distribution}
	\psfrag{T}[l][c]{test set}
	\psfrag{2}[c][c]{2}
	\psfrag{4}[c][c]{4}
	\psfrag{8}[c][c]{8}
	\psfrag{6}[c][c]{6}
	\psfrag{A}[c][c]{relative error [\%]}
	\psfrag{B}[c][c]{$F_{\text{uts}}$}
	\psfrag{D}[c][c]{constant}
	\psfrag{K}[c][c]{width}
	\psfrag{E}[c][c]{constant drill}
	\psfrag{L}[c][c]{hole diameter}
	\psfrag{F}[c][c]{random}
	\psfrag{M}[c][c]{distribution}
	\psfrag{G}[l][c]{initial ANN}
	\psfrag{H}[l][c]{improved ANN}
	\psfrag{20}[c][c]{20}
	\psfrag{40}[c][c]{40}
	\psfrag{50}[c][c]{50}
	\psfrag{10}[c][c]{10}
	\psfrag{30}[c][c]{30}
	\psfrag{0}[c][c]{0}
\includegraphics[width=\textwidth]{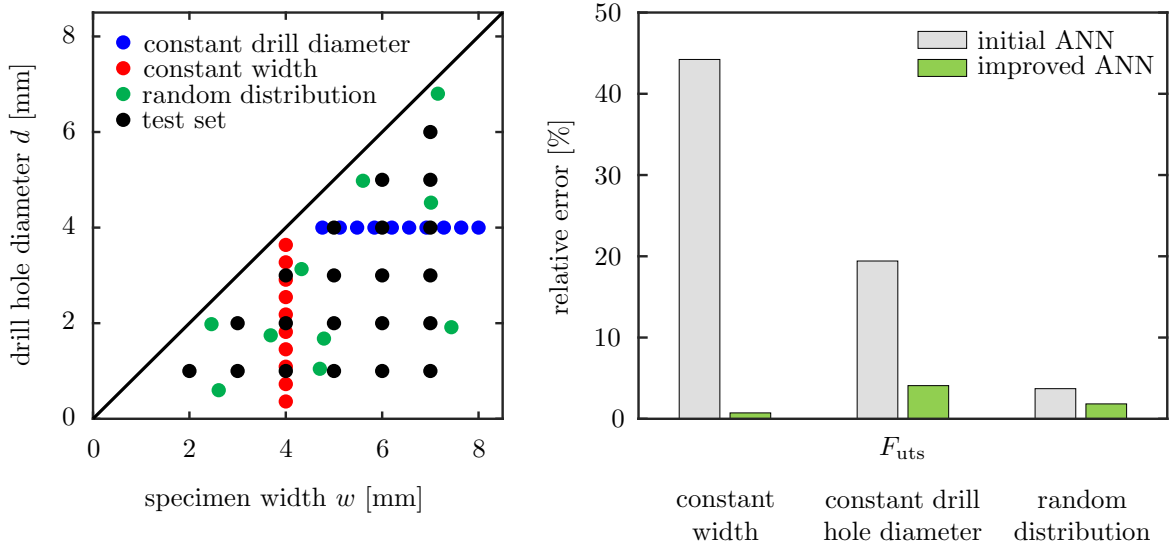}
\caption{Tension of a drilled steelbar: geometry variations for each training data set (left) and relative prediction errors between dimensional ANN and dimensionless (physics-based Rao-Blackwellized) ANN (right).}
\label{fig:drilled_steelbar_example_results}
\end{figure}

The dimensionless ANN$_\text{dim-less}$ performs remarkably better than the dimensional ANN$_\text{dim}$ for all training data sets, see Fig.~\ref{fig:drilled_steelbar_example_results}. Even crude starting predictions are improved by a factor of up to 60 (constant width distribution) and already good starting predictions are further improved by an error reduction of one half (random distribution). The focus on dimensionless sufficient parameters also makes clear what geometric variations should be simulated. A variation of width covers the entire dimensionless domain (which are $d/w$ slopes of the dimensional data set in Fig.~\ref{fig:drilled_steelbar_example_results}) and is to be preferred over a variation of drill hole diameter. Even more so, the physics-based Rao-Blackwellized strategy performs best with less but sufficient simulations, while the dimensional ANN required a larger number of random data points for further improvement. This example demonstrates how a dimensional analysis can be employed to the ANN design to significantly reduce the numerical effort and improve the prediction quality at the same time.

\section{Improvement of ANN output vs.~data vs.~structure:\\ Example of quasi-brittle damage}
Rao-Blackwellization can be applied at different steps of the ANN evolution, though practical operational conditions may limit these options. On the one hand, the ANN structure may be fixed when embedded into a standardized software package or prespecified by convexity or symmetry requirements \citep{klein2022,fernandez2021}. On the other hand, further data acquisition may be unavailable due to confidentiality or experimental limitations. Even intuitive design approaches such as data pre/post-processing may be restricted for various practical reasons. The presented physics-based Rao-Blackwellization scheme must then be adopted to the options of the problem at hand. We will compare three different options in this example, namely, filtering the initial ANN output, augmentation of the training data and adaption of the ANN structure.

To compare the three different options for improvement, we employ ANNs to the modeling of quasi-brittle damage. A common formulation for the energy density of an elastic material undergoing quasi-brittle damage, e.g., for phase-field formulations, depends on strain $\boldsymbol{\varepsilon}$ and damage variable $\alpha$ as \citep{delorenzis2021}
\begin{equation}
W =
\psi(\boldsymbol{\varepsilon}, \alpha)
+ w_1 \left(w(\alpha) + l^2 \|\nabla \alpha\|^2 \right).
\end{equation} 
The first term is the elastic energy density decreasing with damage $\alpha$. The terms inside the brackets correspond to the fracture energy as far as phase-field theory is concerned. In a more general setting, it is responsible for regularization, cf. \citep{langenfeld_2022}. Basic phase-field fracture models however still require an adaption to capture asymmetric tension-compression behavior. We follow the approach of \citet{delorenzis2021} by a decomposition of the elastic energy that also determines the driving force for the damage variable. It takes the form
\begin{align}
\psi(\boldsymbol{\varepsilon}, \alpha) = a(\alpha) \, \psi_\mathrm{D}(\boldsymbol{\varepsilon}) + \psi_\mathrm{R}(\boldsymbol{\varepsilon}) 
\quad
\text{with} 
\quad
 \psi_\mathrm{D}(\boldsymbol{\varepsilon}) + \psi_\mathrm{R}(\boldsymbol{\varepsilon}) = \psi_0(\boldsymbol{\varepsilon}).
\end{align}
Therein, $\psi_\mathrm{R}$ is the residual elastic energy and only $\psi_\mathrm{D}$ is coupled to damage. In the theory of structured deformations \citep{delpiero1993}, the elastic energy of the undamaged material $\psi_0(\boldsymbol{\varepsilon})$ is decreased by the evolution of inelastic deformations $\boldsymbol{\eta}$, the so-called structured deformations. They are constrained to be in a convex set $\mathcal{K}_\varepsilon$ in order to account for the structure of admissible micro-cracks. The elastic energy of the fully damaged body, given $\psi_0(\boldsymbol{\varepsilon})$ and $\mathcal{K}_\varepsilon$, is found by the minimization problem:
\begin{align}
\bar{\boldsymbol{\eta}}(\boldsymbol{\varepsilon}) &= \text{arg min}_{\boldsymbol{\eta} \in \mathcal{K}_\varepsilon}\, \psi_0(\boldsymbol{\varepsilon} - \boldsymbol{\eta}) ,\\
\psi_\mathrm{R}(\boldsymbol{\varepsilon}) &= \text{min}_{\boldsymbol{\eta} \in \mathcal{K}_\varepsilon}\, \psi_0(\boldsymbol{\varepsilon} - \boldsymbol{\eta}) = \psi_0(\boldsymbol{\varepsilon} - \bar{\boldsymbol{\eta}}(\boldsymbol{\varepsilon})).
\end{align}
Accordingly, the idea is similar to that of eigenfracture, cf. \citep{schmidt2009}.
Under certain assumptions, such as a linear elastic and isotropic pristine material behavior, the minimization problem can be specified to
\begin{align}
\bar{\boldsymbol{\eta}}(\boldsymbol{\varepsilon}) = \text{arg min}_{\boldsymbol{\eta}} \left\lbrace \frac{\kappa}{2} \, \text{tr}(\boldsymbol{\varepsilon} - \boldsymbol{\eta})^2 + \mu \| \boldsymbol{\varepsilon}_{\text{dev}} - \boldsymbol{\eta}_{\text{dev}} \|^2, \boldsymbol{\eta} \in \text{Sym: tr}(\boldsymbol{\eta}) \geq \gamma \|\boldsymbol{\eta}_{\text{dev}}\| \right\rbrace , \label{eq:varmin_prob}
\end{align}
wherein $\kappa$ is the bulk modulus, $\mu$ is the shear modulus and $\gamma$ is a free parameter, which controls the ratio between the compressive and the tensile strength in uni-axial loading conditions. The ANNs are employed to predict the resulting energy proportions, $\psi_\mathrm{D}$ and $\psi_\mathrm{R}$. The training data is obtained from the solution provided in \citep{delorenzis2021} and, for better comparison, we assume plane strain-states from now on.  

\subsection{Initial ANN structure for energy prediction}

All of the following ANNs have a network size of 50-50-50-20-2 with a Rectified Linear Unit (ReLu) activation function and are trained for 10000 epochs via the SGD method with a learning rate of 0.01 and a batch size of 1, in order to ensure comparability. Also, five ANNs have been trained for each variant to account for random weight initialization and local minima during training.
The initial ANN takes the strain components $\varepsilon_{11}$, $\varepsilon_{22}$ and $\varepsilon_{12}$ as input parameters. The test data set comprises all possible combinations of the three strain components in the range of $[-0.1 , +0.1]$ with a discretization of $\Delta \varepsilon = 0.001$. Due to the view on practical constraints, three different training data sets with a different amount of data points were generated based on the three strain components, with each component in the range of $[-0.1 , +0.1]$ for comparison. The data distribution is illustrated in Fig. \ref{fig:energy_ann_comparison} (bottom). 

\subsection{Sufficient set from variational minimization problem}
As an additional focus of investigation, this example will highlight the role of sufficiency. Three types of sufficient information will be distinguished for the variant of structure improvement. Firstly, non-sufficient information lacks some part of the sufficient set. Secondly, sufficient information fully determines the unknown parameter. Thirdly, minimally sufficient information is not only sufficient but cannot be reduced further without becoming non-sufficient.

From the minimization problem \eqref{eq:varmin_prob} one can conclude that one sufficient set is given by
\begin{align}
\mathcal{S} &= \{ \left\Vert \boldsymbol{\varepsilon}_\mathrm{dev} \right\Vert, \text{tr}( \boldsymbol \varepsilon ) \}.
\end{align}
Fixing the deviatoric norm and the trace of the strains will determine the evolution of the inelastic deformations and finally the individual energy contributions $\psi_\mathrm{R}$ and $\psi_\mathrm{D}$. Again, we do not need to know how this relationship looks like.
This moreover allows to specify a non-sufficient, a minimal sufficient and a non-minimal sufficient set, reading
\begin{align}
\mathcal{S}_1 &= \{ \left\Vert \boldsymbol{\varepsilon}_\mathrm{dev} \right\Vert \} && \text{(non-sufficient)}, \\
\mathcal{S}_2 &= \{ \left\Vert \boldsymbol{\varepsilon}_\mathrm{dev} \right\Vert, \text{tr}( \boldsymbol \varepsilon ) \} && \text{(minimal sufficient)},\\
\mathcal{S}_3 &= \{ \left\Vert \boldsymbol{\varepsilon}_\mathrm{dev} \right\Vert, \text{tr}( \boldsymbol \varepsilon ), \varepsilon_{11} \} && \text{(non-minimal sufficient)}.
\end{align}
Note that the initial ANN input is also a non-minimal sufficient set $\mathcal{S}_4 =\{ \varepsilon_{11},\varepsilon_{22},\varepsilon_{12} \}$.

\subsection{Improved output, data and structure in comparison}
\begin{figure}[h]
\centering
	\psfrag{A}[c][c]{mean squared error}
	\psfrag{B}[l][c]{initial ANN}
	\psfrag{G}[l][c]{improved ANN (output filter by $\mathcal{S}_2$ - minimal sufficient)}
	\psfrag{C}[l][c]{improved ANN (data augmentation by $\mathcal{S}_2$ - minimal sufficient)}
	\psfrag{D}[l][c]{improved ANN (structure-optimized by $\mathcal{S}_2$ - minimal sufficient)}
	\psfrag{E}[l][c]{\phantom{improved ANN}\, (structure-optimized by $\mathcal{S}_1$ - non-sufficient)}
	\psfrag{F}[l][c]{\phantom{improved ANN}\, (structure-optimized by $\mathcal{S}_3$ - non-minimal sufficient)}
	\psfrag{10}[l][c]{$10^{-6}$}
	\psfrag{20}[l][c]{$10^{-5}$}
	\psfrag{30}[l][c]{$10^{-4}$}
	\psfrag{40}[l][c]{$10^{-3}$}
	\psfrag{50}[l][c]{$10^{-2}$}
	\psfrag{60}[l][c]{$10^{-1}$}
	\psfrag{0}[c][c]{0}
	\psfrag{0.1}[c][c]{0.1}
	\psfrag{-0.1}[c][c]{-0.1}
	\psfrag{x}[c][c]{$\varepsilon_{11}$}
	\psfrag{y}[c][c]{$\varepsilon_{22}$}
	\psfrag{z}[c][c]{$\varepsilon_{12}$}
\includegraphics[width=0.9\textwidth]{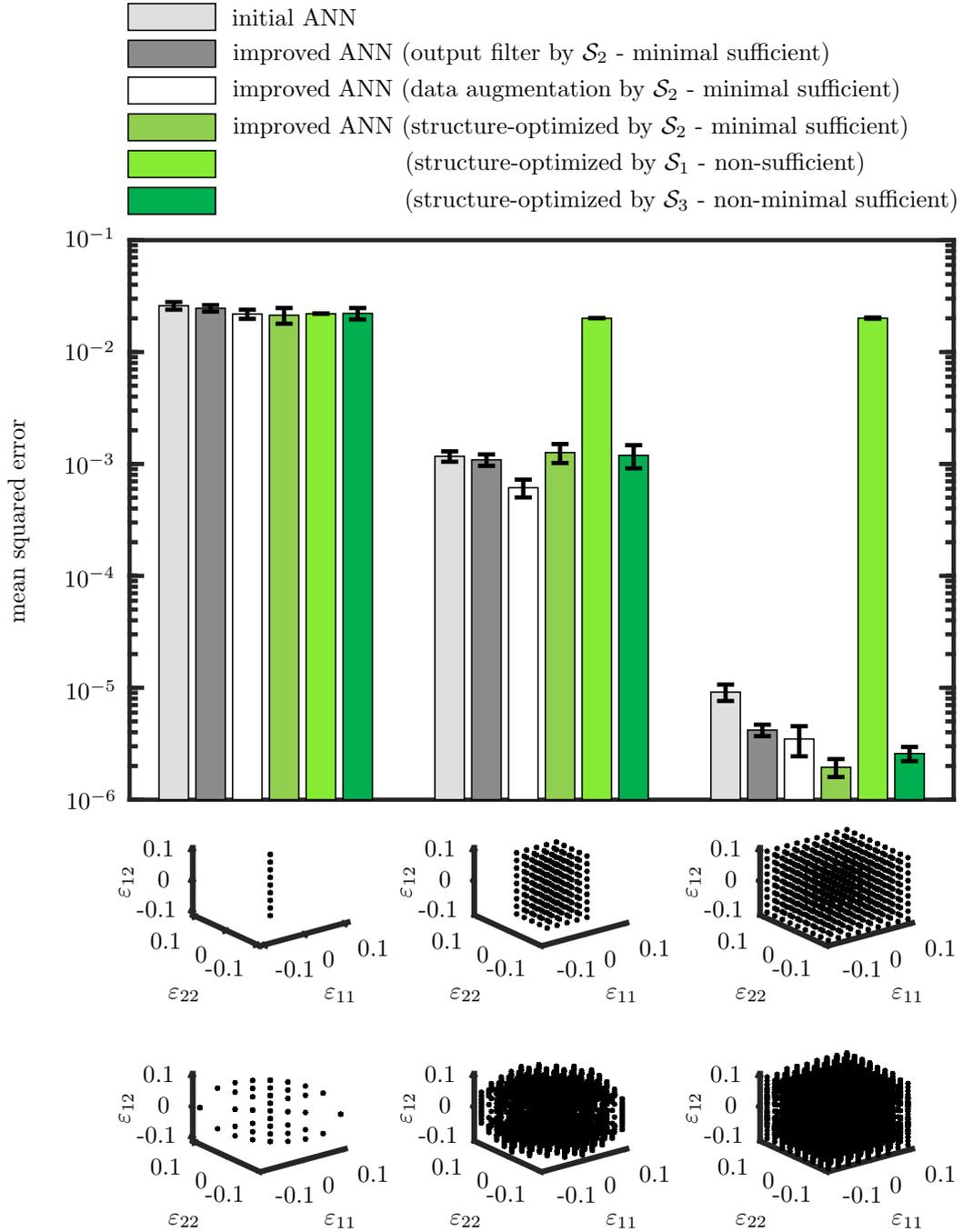}
\caption{Comparison of mean squared errors of the initial, output filtered, data-augmented as well as structure-optimized ANNs. Standard deviations are provided based on five, randomly assigned initial weight distributions. Corresponding training data sets for the initial and structure-optimized ANNs are depicted in the upper row, while the ones of the data augmentation are shown in the lower row.}
\label{fig:energy_ann_comparison}
\end{figure}

Filtering the output and training data augmentation always yields to improvements in terms of the mean squared error, with maximum factors of 2.2 (output filtering) and 2.6 (data augmentation) for the largest training data set, see Fig. \ref{fig:energy_ann_comparison} (dark gray and white bars compared to light gray bars). Throughout all training data sets, training-data augmentation leads to better results compared to the output filter method. The observed behavior can be explained by the fact that the data augmentation method extrapolates the training data values in the strain component space, thereby covering more of the test data space. Data augmentation is only outperformed by structure optimization of the first layer for the smallest and largest data. For the largest training data set, which has the same parameter range as the test data, the structure-optimized ANN shows a 4.7 times lower error compared to the initial and a 1.8 times lower error compared to the data augmentation ANN.

Two practical limitations emerge along with the generally favorable results of the physics-based Rao-Blackwell scheme. Firstly, the smallest data set is simply too small in order to represent the test data properly. It leads to unsatisfactory predictions for all strategies, despite the observed improvement. Secondly, the structure-optimized ANN trained on the medium-sized data set performs even worse than the initial ANN. This usually contradicts the error inequality of the analytical definition, but the slight deviation can be explained by a poorer numerical extrapolation and possible local minima. Another reason worth mentioning is that the initial input is already a non-minimal sufficient set such that an improvement can not be guaranteed here.

The comparison of different kinds of sufficiency shows a poor performance for the non-sufficient set ($\mathcal{S}_1$). This result can be expected, but it also underlines quantitatively how well the sufficient set performs compared to a non-sufficient one. For the non-minimal sufficient set $\mathcal{S}_3$, the error is similar to that of the minimal sufficient set, except for the largest training data set, where the error is slightly higher. One possible explanation for this increased error is that the ANN cannot fully eliminate the influence of the dispensable parameter $\varepsilon_{11}$. The presented employment of physics-based Rao-Blackwellization thus clearly helps to efficiently use knowledge about sufficient, but not superfluous, physical information.

In conclusion for the present example of damage energy prediction, physics-based Rao-Black\-wellization should be applied via structure optimization if the training data set covers the entire parameter range. It should be used for data augmentation instead, if the training data set covers a smaller parameter range than required by the intended application. Without prior indication, the changes of the proposed scheme range between numerically negligible deviations and clear improvements.


\clearpage
\section{When physical information can improve ANNs and when it cannot: Experimental example of isotropic elasticity}

This example resumes the introductory motivation in Fig.~\ref{fig:introductory_example_poisson_ratio}, where a rash implementation of physical constraints was shown to worsen a prediction. We now expand the introductory case to a complete ANN prediction of isotropic rubber elasticity trained on DIC data of a highly-elastic 3D-printed TPU (Thermoplastic Polyurethane, Ultimaker 95A) specimen under plane-stress conditions. The principle of utilizing isotropy or homogeneity is already well-known from analytical invariant formulations as well as data-driven predictions, e.g., for PINNs and CANNs \citep{tang2019,raissi2019,linka2021}. Here, we will analyze how the employment of isotropy, homogeneity and differentiability follows naturally from physics-based Rao-Blackwellization. At the same time, we will show that the improper employment of auxetic and incompressible limits is not advocated by this strategy and how the modifications perform when comparing numerical predictions with original experimental data.

\subsection{Initial data set}
The initial state is determined by the raw strain and stress data obtained from tensile stretch experiments (in a Kammrath \& Weiss tensile module) with simultaneous picture recording for DIC (via Veddac Strain software), see Fig.~\ref{fig:isoann_comp}. Data is recorded for increasing steps of clamp displacement. Piola-Kirchhoff stress values are derived from corresponding force measurements and initial cross-sectional area. Strain data is determined from DIC for 275 subregions evenly distributed in the central region of the specimen.

Considering analytical model descriptions, this data could be used to fit a hyperelastic isotropic model. Application to an ANN instead requires further adaption. The stress and strain data does not cover the entire domains, e.g., in terms of other load directions. An ANN trained only on the raw data will thus not only perform poorly. It will simply fail, e.g., already prohibiting convergence as part of a finite element framework. Among multiple options to adapt the ANN (e.g., see the previous examples improving ANNs in terms of training data, structure or output filters), this example will apply physics-based Rao-Blackwellization directly to the experimental data to better discuss the assumptions employed to the model.

\subsection{Data augmentation by sufficient and other physical knowledge}
A sufficient information set depends on the respective model (both physical or statistical) and can only be obtained by knowledge about the underlying assumptions. 
To prepare the improvement step of physics-based Rao-Blackwellization, we want to examine six constraints for the further discussion:

\begin{enumerate}
\item A homogeneous stress state sufficiently determines a homogeneous strain state and vice versa for each deformation state,
\item the isotropic material dictates material invariance under rotations,
\item the stress-strain relations are continuously differentiable,
\item the loading state remains plane-stress,
\item the material is not incompressible,
\item the material is not auxetic.
\end{enumerate}

It should be emphasized that modeling assumptions naturally show deviations from the real system, some of which can be obvious and some less obvious. In the present example, the force measurements are almost tacitly assigned to stress values inside the central region of the specimen. Strains are however recorded via DIC and show heterogeneous fluctuations that can arise from material imperfections but also from the evaluation algorithm and limited pixel resolution, for instance. Such incoherent data structure can pose a practical challenge for the questions raised in the motivation: what information to incorporate, when to do it and how to do it correctly? Physics-based Rao-Blackwellization provides a reliable answer for those assumptions that yield sufficient information.

\subsection{Improved ANN material models and incompatible constraints}

\subsubsection{Employing model assumptions}
The first model assumption of the above list implies that the (assumed) homogeneous stress state must fully determine the (equally assumed homogeneous) strain state in the central region of the specimen. The specimen shape was accordingly chosen to provide a homogeneous field until the onset of necking, which is not assumed nor observed for the present load regimes. Fluctuations of the DIC-based strain data nevertheless occur and are treated by physics-based Rao-Blackwellization. It dictates for every deformation step, that the sufficient stress value is assigned to a strain value determined from the conditionally averaged-based estimator as defined in \eqref{eq:theta1}. This yields to a unique stress-strain path, for example, cf. Fig.~\ref{fig:isoann_uniaxial_tension}. 

The second model assumption ensures isotropic material behavior and can be used for data augmentation, because stresses and strains sufficiently determine each other, modulo rotations. In the sense of physics-based Rao-Blackwellization, new stress-strain data pairs $\{\boldsymbol{\sigma}^\star,  \boldsymbol{\varepsilon}^\star \}$ are thus generated by rotating a measured pair $\{\boldsymbol{\sigma},  \boldsymbol{\varepsilon} \}$ by rotation matrix $\boldsymbol{R}$ as 
\begin{align}
\boldsymbol{\varepsilon}^\star = \boldsymbol{R}^T(\vartheta) \cdot \boldsymbol{\varepsilon} \cdot \boldsymbol{R}(\vartheta),\\
\boldsymbol{\sigma}^\star = \boldsymbol{R}^T(\vartheta) \cdot \boldsymbol{\sigma} \cdot \boldsymbol{R}(\vartheta).
\end{align}
The rotation angle $\vartheta$ ranges from $0$ to $2\pi - \Delta \vartheta$ in discrete steps of $\Delta \vartheta = \pi/18$. 

The third model assumption does not only guarantee a continuous tangent stiffness. Physics-based Rao-Blackwellization implies that the elastic tangent in the tension regime sufficiently also determines the behavior under compression around the origin. Compression states are not substantially triggered, neither for training data nor for its evaluation. They are thus obtained, up to 1.016\,\% compression strain, by flipping tension states via linear regression. This expansion remains in the linear regime and fully covers the compressive states that appear during the simulation tests.

The fourth model assumption allows to neglect out-of-plane contributions of the stress tensor as information. Accordingly, transverse contraction in the third direction is implicitly accounted for by the ANN but not resolved explicitly.

The fifth and sixth model assumptions of the above list cannot be applied to the original data in the present context or physics-based Rao-Blackwellization, because they cannot be utilized for sufficient information transfer. This should of course not be understood as an evaluation of their importance or relevance. These two constraints are physically sound and should be considered for the solution of mechanical problems with standard materials. Their incompatibility yet indicates that their incorporation into the Rao-Blackwell scheme cannot guarantee an improvement, which was already indicated by the introductory example and will be examined quantitatively now. 

A combination of all the above constraints is hence not suggested by physics-based Rao-Blackwell\-ization. We will thus compare two improvements of the data set for ANN training. The first data set is improved by constraints 1-4 and shall simply be labeled as "RB". The second data set additionally considers constraints 5 and 6 by removing unphysical data and shall be labeled as "RB+inc+aux". 

\subsubsection{ANN-based finite element simulation compared to experimental reference}
\begin{figure}[h]
\centering
	\psfrag{A}[c][c]{Experiment and DIC}
	\psfrag{B}[c][c]{ANN RB}
	\psfrag{C}[c][c]{ANN RB+inc+aux}
	\psfrag{D}[c][c]{$\varepsilon_{\text{xx}}$}
	\psfrag{E}[c][c]{$\varepsilon_{\text{yy}}$}
	\psfrag{F}[c][c]{reference configuration}
	\psfrag{N}[c][c]{boundary value problem}
	\psfrag{R}[c][c]{\color{red}{$u/2$}}
	\psfrag{G}[c][c]{$\sigma_{\text{xx}}$}
	\psfrag{J}[c][c]{$\sigma_{\text{yy}}$}
	\psfrag{a}[l][c]{0}
	\psfrag{n}[l][c]{0}
	\psfrag{e}[l][c]{0.4}
	\psfrag{f}[l][c]{-0.4}
	\psfrag{b}[l][c]{0.15}
	\psfrag{c}[l][c]{-0.15}
	\psfrag{d}[l][c]{-0.3}
	\psfrag{g}[l][c]{0.02}
	\psfrag{k}[l][c]{0.0075}
	\psfrag{h}[l][c]{0.0138}
	\psfrag{x}[l][c]{0.005}
	\psfrag{p}[l][c]{-0.01}
	\psfrag{y}[l][c]{-0.05}
	\psfrag{o}[l][c]{-0.005}
	\psfrag{q}[l][c]{0.75}
	\psfrag{r}[l][c]{0.5}
	\psfrag{s}[l][c]{0.25}
	\psfrag{u}[l][c]{0.2}
	\psfrag{v}[l][c]{0.075}
	\psfrag{P}[c][c]{$10$ mm}
	\psfrag{T}[c][c]{$20$ mm}
	\psfrag{3.8}[c][c]{3.8 mm}
	\psfrag{33.8}[c][c]{33.8 mm}
	\psfrag{20.8}[c][c]{20.8 mm}
	\psfrag{19.9}[c][c]{19.9 mm}
\includegraphics[width=0.95\textwidth]{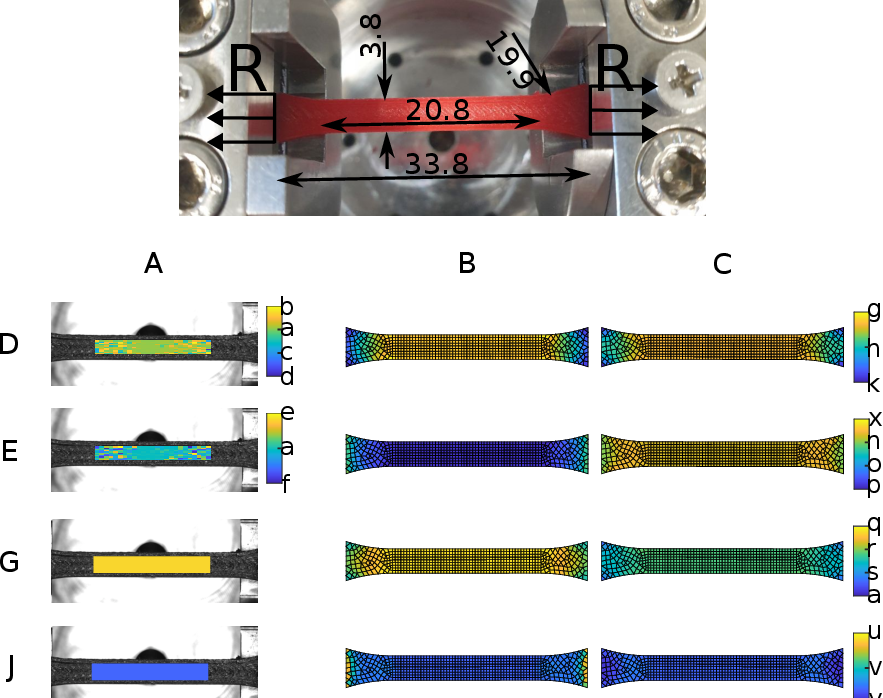}
\caption{Experimental set-up (top) and comparison of stress and strain fields determined in  the experiment as well as in the finite element simulations for $u = 0.577$ mm clamp displacement (bottom). Experimental strain fields are obtained from DIC and stress values are calculated from force and area measurements with free lateral boundary conditions, respectively.
}
\label{fig:isoann_comp}
\end{figure}

The improved data sets are used to train an ANN, "RB" and "RB+inc+aux" respectively, which serves the role of a material model routine in finite element simulations. Note that the raw data set would not yield any feasible result due to the severe lack of training data in the stress and strain domains. The crude initial estimator, in the sense of physics-based Rao-Blackwellization, thus simply fails already by not being able to perform a practicable finite element simulation. The ANNs have a network size of 10-10-10-10-3 with a ReLu activation function and are trained via the SGD method for various epochs with adaptive learning rates and batch sizes. The training process yields functional ANNs on the material point level for the finite element simulations. The original experimental setting is finally simulated for evaluation of the ANN performance. Therefore, the specimen is meshed via triangular and quadrilateral elements with linear ansatz functions. Boundary conditions are applied according to the depicted boundary value problem, see Fig.~\ref{fig:isoann_comp}.

Interestingly, both ANNs yield very similar longitudinal strain ($\varepsilon_{xx}$) predictions. These also match the experimental results well. This can be explained by the fact that lateral strain was already well captured during the DIC evaluation and has not been significantly changed by modeling assumptions such as isotropy or differentiability. Also the lateral stress in the central specimen region is similar between the ANNs. Both adopted well to the stress-free boundary condition that is also already represented by the original data. A first noticeable difference appears at the clamped ends, though. The "RB+inc+aux" ANN, additionally disregarding incompressible and auxetic limits, predicts neutral to compressive instead of more realistic tensile conditions. 

\begin{figure}[h]
\centering
	\psfrag{C}[l][c]{Strain-averaged experimental data (model assumption 1)}
	\psfrag{D}[l][c]{ANN RB}
	\psfrag{E}[l][c]{ANN RB+inc+aux}
	\psfrag{F}[l][c]{Compression data (model assumption 3)}
	\psfrag{G}[l][c]{Raw experimental data}
	\psfrag{A}[c][c]{$\varepsilon_{\text{xx}}$ [-]}
	\psfrag{B}[c][c]{$\sigma_{\text{xx}}$ [MPa]}
	\psfrag{0}[c][c]{0}
	\psfrag{0.02}[c][c]{0.02}
	\psfrag{0.2}[c][c]{0.2}
	\psfrag{-0.2}[c][c]{-0.2}
	\psfrag{-0.3}[c][c]{-0.3}
	\psfrag{-0.1}[c][c]{-0.1}
	\psfrag{0.05}[c][c]{0.05}
	\psfrag{-0.05}[c][c]{-0.05}
	\psfrag{0.04}[c][c]{0.04}
	\psfrag{0.06}[c][c]{0.06}
	\psfrag{0.08}[c][c]{0.08}
	\psfrag{0.1}[c][c]{0.1}
	\psfrag{0.4}[c][c]{0.4}
	\psfrag{0.6}[c][c]{0.6}
	\psfrag{0.8}[c][c]{0.8}
	\psfrag{1}[c][c]{1}

\includegraphics[width=0.75\textwidth]{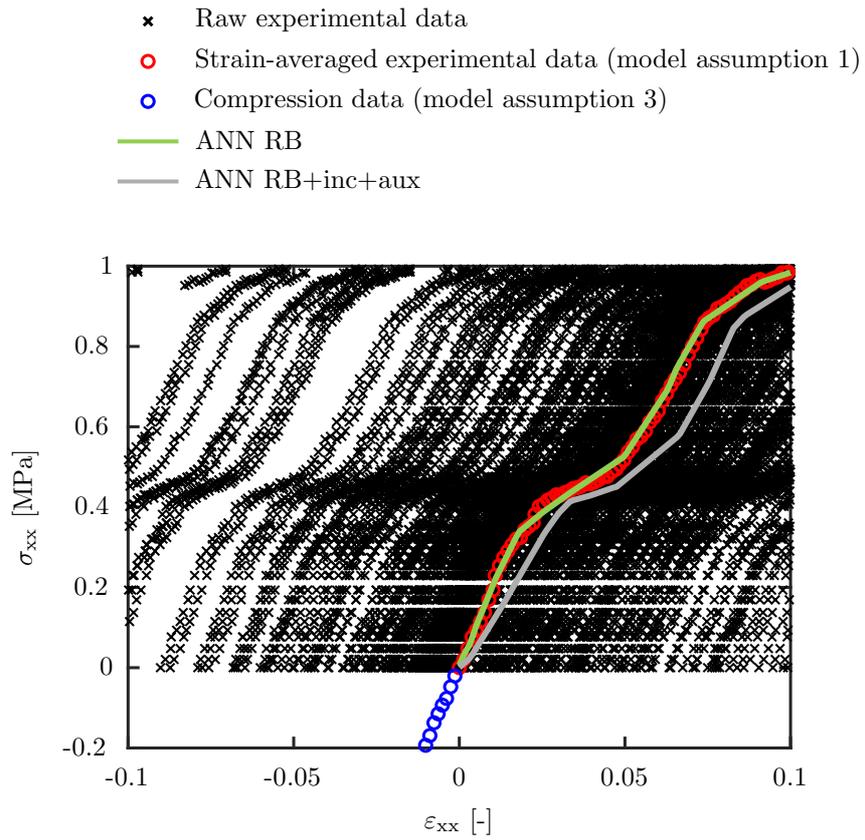}
\caption{Raw stress-strain data under uniaxial tension from experimental data and application of assumptions 1 (homogeneity) and 3 (differentiability) and curves predicted by the ANNs trained with different model assumptions, RB vs RB+inc+aux. 
}
\label{fig:isoann_uniaxial_tension}
\end{figure}

The strongest deviation between both ANNs, "RB" and "RB+inc+aux", appears for longitudinal stress and lateral strain. The first is already indicated by the stress-strain curve in Fig.~\ref{fig:isoann_uniaxial_tension}. Examining the raw data in detail, indicates a systematic error due to limiting DIC resolution. It shifts the strain origin for individual pixels while maintaining the overall course correctly. But again, we want to mimic a scenario that does not allow a retrospective correction of the experimental setup and focus in the present work on improvement with physics-based information instead. While the qualitative evolution is still captured by both ANNs, only the "RB" case based on clean physics-based Rao-Blackwellization allows for a reliable quantitative prediction of the stress-strain relationships. Omitting incompressible and auxetic states almost halves the elastic tangent in some strain states, caused be the strain bias also already discussed in the introduction. The lateral strain shows another severe difference between the ANNs, because insensitive training data becomes more relevant and depends even more on reliable data processing, cf.~Fig.~\ref{fig:isoann_comp}. To be more specific in terms of the $-\varepsilon_{\text{yy}} / \varepsilon_{\text{xx}}$ ratio, the "RB+inc+aux" case shows a deviation of 119\,\% from the experimental data for 0.577 mm clamp displacement. The deviation of the pure improvement case "RB" is only 7.25\,\% and hence 16 times smaller. 

Eventually, these results show that the physics-based Rao-Blackwellization can be successfully used with model assumptions that allow to transfer sufficient information into a finite-element framework. The original experiment is numerically well captured when relying on the pure improvement algorithm. It cannot incorporate all assumptions such as incompressible and auxetic limits, yet it clearly indicates whether such information is not suitable and should be left out --- guaranteeing improvement even for crude initial estimators or data availability.

%

\clearpage
\section{Conclusion}

The concept of Rao-Blackwellization was adapted from its statistical origin to deterministic physical problems. It has been successfully used to improve ANN predictions in the context of material modeling. The basic idea is to unify an ANN's prediction over sufficient information sets. Identifying sufficient information is the key input for this strategy. It can be obtained from various sources for the problem at hand, for instance: parameters of governing equations fixing the solution, fundamental laws, assumptions like isotropic or quasi-static problems and dimensional analysis. The improvement scheme can then be employed into the ANN by output filtering, training data augmentation or structure adaption.

Complementing existing approaches to incorporate physical knowledge, the proposed strategy provides an error norm and a guarantee in the form of an error inequality. The presented examples showed various opportunities for the improvement of ANNs. The classification example of an elasto-plastic predictor demonstrated how the scheme suppresses noise and overfitting by post-processing of the output. The simulations of the drilled steel bar modified the role of the ANN in the engineering problem of force predictions. Outsourcing the dimensional scaling makes the ANN part much more efficient and less demanding in terms of data points and demanding numerical resources. The example of quasi-brittle damage underlined why sufficient information is required for guaranteed improvement and compared output filtering to data augmentation and structure optimization. Finally, the DIC measurements of flexible, 3D-printed TPU demonstrated that isotropy, homogeneity and differentiability can support the ANN training. The proposed strategy can thus support intuitive as well as less intuitive modeling assumptions. It was shown at the same time, though, that even sound physical constraints can lead to poorer predictions if they are not supported by a clear improvement scheme such as Rao-Blackwellization. 

The present strategy thus complements the scope of existing approaches such as PINNs or CANNs by focusing on the information used, not on the form of the solution. As a first relevant feature, it is guaranteed not to worsen the initial prediction. Secondly, it can improve even crude initial estimators and is thus suitable for situations with an incomplete data basis. Thirdly and more practically, it can be applied flexibly to data, structure or integration of the ANN.

These results shall also promote further exchange between statistical modeling and material modeling. Follow-up studies will aim at mathematical aspects such as the impact of parameterization on individual error weighting and numerical inaccuracies. 
Future work will moreover apply the scheme to information transfer in multi-scale problems, for example, obtaining sufficient information from homogenization of mechanial systems and scale separation \citep{vrugt2021,javili2017,heitbreder2018,kurzeja2016,cushman2002}.

\begin{appendix}
\section{Proving how the new prediction never gets worse}\label{chp_math_proof}

The following proof is not fully rigorous in a mathematical sense as some properties are tacitly assumed. Most importantly, we assume that there is a relationship (e.g., between stress and yield behavior) and that integrals are well defined. More information about such peculiarities can be found in \citep{blackwell1947, lehmann1998, rao1973, bickel2006}. These assumptions are however weak when considering well-defined material models with finite deformation energies, limited deformation ranges and rates. The key idea of the following proof closely follows the original Rao-Blackwell theorem and is based on a split of the parameterization into sufficient parameters and remaining parameters not affecting the unknown of interest.

Suppose the correct model prediction is given by $\theta_\star$, then the error of the initial prediction is defined as
\begin{align}
\delta_0 = \frac{1}{\vert \Omega \vert} \int\limits_\Omega (\theta_0 - \theta_\star)^2 \, \diff \omega. \label{eq:err0}
\end{align}
This can be rewritten as 
\begin{align}
\frac{1}{\vert \Omega \vert} \int\limits_\Omega (\theta_0 - \theta_\star)^2 \, \diff \omega = \frac{1}{\vert \Omega \vert} \int\limits_\Omega (\theta_0 - \theta_1 + \theta_1 - \theta_\star)^2 \, \diff \omega
\end{align}
using the improved model $\theta_1$ and with the binomial theorem
\begin{align}
\frac{1}{\vert \Omega \vert} \int\limits_\Omega (\theta_0 - \theta_\star)^2 \, \diff \omega = &\frac{1}{\vert \Omega \vert} \int\limits_\Omega (\theta_0 - \theta_1)^2 \, \diff \omega + \frac{1}{\vert \Omega \vert} \int\limits_\Omega (\theta_1 - \theta_\star)^2 \, \diff \omega \label{eq:err0re} \\
 &+ \frac{2}{\vert \Omega \vert} \int\limits_\Omega (\theta_0 - \theta_1) \, (\theta_1 - \theta_\star) \, \diff \omega \nonumber 
\end{align}
follows. To show that the last term of the right-hand side of \eqref{eq:err0re} vanishes, the integral over the domain $\Omega$ is split into two parts. One integrating over the domain $\Omega_s$, where $\mathcal{S}(\omega)=s$ is constant, and another integrating over the remaining parameterization of the physical state $\Omega_t$. This leads to
\begin{align}
\frac{1}{\vert \Omega \vert} \int\limits_\Omega (\theta_0 - \theta_1) \, (\theta_1 - \theta_\star) \, \diff \omega = \frac{1}{\vert \Omega_t \vert} \int\limits_{\Omega_{t}} \frac{1}{\vert \Omega_s \vert} \int\limits_{\Omega_{s}} (\theta_0 - \theta_1) \, (\theta_1 - \theta_\star) \, \diff \omega \diff s\label{eq:zeroterm}
\end{align}
and, because $\theta_1$ is constant in $\Omega_s$,
\begin{align}
\frac{1}{\vert \Omega \vert} \int\limits_\Omega (\theta_0 - \theta_1) \, (\theta_1 - \theta_\star) \, \diff \omega = \frac{1}{\vert \Omega_t \vert} \int\limits_{\Omega_{t}} (\theta_1 - \theta_\star) \, \frac{1}{\vert \Omega_s \vert} \int\limits_{\Omega_{s}} (\theta_0 - \theta_1) \, \diff \omega \diff s. \label{eq:zerotermre}
\end{align}
Considering only the inner integral and splitting it up, one obtains 
\begin{align}
\frac{1}{\vert \Omega_s \vert} \int\limits_{\Omega_{s}} (\theta_0 - \theta_1) \, \diff \omega = \frac{1}{\vert \Omega_s \vert} \int\limits_{\Omega_{s}} \theta_0 \, \diff \omega - \frac{1}{\vert \Omega_s \vert} \int\limits_{\Omega_{s}} \theta_1 \, \diff \omega.
\end{align}
This integral vanishes, since
\begin{align}
\frac{1}{\vert \Omega_s \vert} \int\limits_{\Omega_{s}} \theta_0 \, \diff \omega - \frac{\theta_1}{\vert \Omega_s \vert} \int\limits_{\Omega_{s}} 1 \, \diff \omega = \theta_1 - \theta_1 = 0 \label{eq:zeroconfirm}
\end{align}
by definition of $\theta_1$. Now, resubstitution of \eqref{eq:zeroconfirm} in equations \ref{eq:zerotermre} and \ref{eq:zeroterm} and solving \eqref{eq:err0re} for the error of the improved model $\theta_1$ finally leads to
\begin{align}
\frac{1}{\vert \Omega \vert} \int\limits_\Omega (\theta_1 - \theta_\star)^2 \, \diff \omega = \frac{1}{\vert \Omega \vert} \int\limits_\Omega (\theta_0 - \theta_\star)^2 \, \diff \omega - \underbrace{\frac{1}{\vert \Omega \vert} \int\limits_\Omega (\theta_0 - \theta_1)^2 \, \diff \omega}_{\geq 0}.
\end{align}
Since the second term of the right-hand side is always greater or equal to zero, it is confirmed that $\theta_1$ results in a lower or equal error compared to the one of the initial model $\theta_0$
\begin{align}
\frac{1}{\vert \Omega \vert} \int\limits_\Omega (\theta_1 - \theta_\star)^2 \, \diff \omega \leq \frac{1}{\vert \Omega \vert} \int\limits_\Omega (\theta_0 - \theta_\star)^2 \, \diff \omega
\end{align}
and the proof is complete. Also note that this scheme is idempotent and another Rao-Blackwellization of $\theta_1$ does not lead to further improvement $\theta_2$, since
\begin{align}
\theta_2 = \mathbb{A}\left(\theta_1 \, \vert \, \mathcal{S} = s\right) = \frac{1}{\vert \Omega_s \vert} \int\limits_{\Omega_s} \theta_1 \, \diff \omega = \theta_1.
\end{align}

\section{Overview of ANN hyperparameters} \label{chp_appendix}
The following table summarizes the ANN hyperparameters for every example. They were chosen according to convergence studies and by taking into account problem specific requirements. For the elasto-plastic predictor example the Tanh delivers a suitable activation function due to its bounded output. For all other examples the common ReLu function provided the best result when combined with a linear activation of the last layer. 
\begin{table}[h!]
\resizebox{\textwidth}{!}{%
\begin{tabular}{|l|c|c|c|c|c|}
\hline
Example                                                            & Network size                                                        & \begin{tabular}[c]{@{}c@{}}Activation\\ function\end{tabular} & \begin{tabular}[c]{@{}c@{}}Training\\ algorithm\end{tabular} & Learning rate & Batch size \\ \hline
\begin{tabular}[c]{@{}l@{}}Elasto-plastic\\ predictor\end{tabular} & \begin{tabular}[c]{@{}c@{}}5-1\\ 10-5-1\\ 200-200-60-1\end{tabular} & Tanh                                                          & SGD                                                          & adaptive      & adaptive   \\ \hline
Drilled steel bars                                                 & 13-13-13-1                                                          & ReLu*                                                          & SGD/Adam                                                      & adaptive      & adaptive   \\ \hline
\begin{tabular}[c]{@{}l@{}}Quasi-brittle\\ damage\end{tabular}     & 50-50-50-20-2                                                       & ReLu*                                                          & SGD                                                          & 0.01          & 1          \\ \hline
Isotropic elasticity                                               & 10-10-10-10-3                                                       & ReLu*                                                          & SGD                                                          & adaptive      & adaptive   \\ \hline
\end{tabular}%
}
\caption{ANN hyperparameters for the different examples.\\
*last layer has a linear activation function}
\end{table}
\end{appendix}

\bibliographystyle{apalike-ejor}%
\bibliography{Literatur}%
\end{document}